\documentclass[journal]{IEEEtran}
\usepackage{graphicx} 
\usepackage{array} 
\usepackage{multirow}
\usepackage{subfigure} 
\usepackage{float}
\usepackage{amsmath}
\usepackage{amsthm}
\usepackage{amssymb}
\usepackage{threeparttable}
\usepackage{balance}
\usepackage{amsmath,bm}
\usepackage{wasysym}
\usepackage[linesnumbered,ruled,vlined]{algorithm2e}
\usepackage{amsfonts}
\usepackage{booktabs}
\usepackage[table]{xcolor}
\usepackage{blindtext}
\usepackage[T1]{fontenc}
\usepackage[utf8]{inputenc}
\usepackage{cleveref}
\crefname{section}{§}{§§}
\Crefname{section}{§}{§§}
\SetKwProg{Fn}{Function}{}{}

\newcommand{\nop}[1]{}
\usepackage{xcolor}

\newcommand{\tabincell}[2]{\begin{tabular}{@{}#1@{}}#2\end{tabular}}

\nop{}

\newtheorem{remark}{Remark}

\begin{document}
\title{PLVER: Joint Stable Allocation and Content Replication for Edge-assisted Live Video Delivery}

\author{Huan~Wang,
        Guoming~Tang,~\IEEEmembership{Member,~IEEE,}
        Kui~Wu,~\IEEEmembership{Senior~Member,~IEEE,}
        and~Jianping~Wang~\IEEEmembership{Member,~IEEE}
\thanks{H. Wang and K. Wu are with the Department of Computer Science, University of Victoria, Victoria, BC V8W 3P6, Canada (e-mail: \{huanwang, wkui\}@uvic.ca).}
\thanks{G. Tang is with Peng Cheng Laboratory, Shenzhen, Guangdong 518066, China (e-mail: tanggm@pcl.ac.cn).}
\thanks{J. Wang is with Department of Computer Science, City University of Hong Kong, Hong Kong (e-mail: jianwang@cityu.edu.hk).}
}

\maketitle

\begin{abstract}
The live streaming services have gained extreme popularity in recent years. Due to the spiky traffic patterns of live videos, utilizing the distributed edge servers to improve viewers' quality of experience (QoE) has become a common practice nowadays. Nevertheless, current client-driven content caching mechanism does not support caching beforehand from the cloud to the edge, resulting in considerable cache missing in live video delivery. State-of-the-art research generally sacrifices the liveness of delivered videos in order to deal with the above problem. In this paper, by jointly considering the features of live videos and edge servers, we propose \textit{PLVER}, a proactive live video push scheme to resolve the cache miss problem in live video delivery. Specifically, \textit{PLVER} first conducts a one-to-multiple stable allocation between edge clusters and user groups, to balance the load of live traffic over the edge servers. Then it adopts proactive video replication algorithms to speed up the video replication among the edge servers. We conduct extensive trace-driven evaluations, covering $0.3$ million Twitch viewers and more than $300$ Twitch channels. The results demonstrate that with \textit{PLVER}, edge servers can carry $28\%$ and $82\%$ more traffic than the auction-based replication method and the caching on requested time method, respectively.

\end{abstract}

\nop{
The edge cache miss problem exists for live video is mainly due to two reasons: i) different from the requests of regular contents, too many simultaneous live video requests arrives the edge too fast even before the caching process is finished at the edge layer, and ii) the edge caching process in the current video delivery architecture is normally triggered by the client requests. In other words, the video segment caching (replication) process (from cloud to the edge) in the current architecture will only commence when the cloud receives the first request for that segment (as shown in Fig.~\ref{fig:motivation_gap}). While this strategy makes sense when delivering regular content, it slows down the caching process in the context of live videos: there exists a time \textit{gap} (as shown in Fig.~\ref{fig:motivation_gap}) between the time when a segment is generated from the cloud and when the cloud receives the first request. }

\section{Introduction}
The last few years have witnessed the dramatic proliferation of live video streams over streaming platforms (such as Twitch, Facebook
Live, and Youtube Live, etc.) which have generated billion dollars of revenue~\cite{wang2019intelligent}. According to the statistics of Twitch, in 2019, over $660$ billion minutes of live streams were watched by customers and $3.64$ million streamers (monthly average) broadcast their channels via Twitch~\cite{twitch2019statis}. 

Nevertheless, the delivery of live videos is quite different from the conventional video-on-demand (VoD) service. First, live video has quite spiky traffic, which means the viewer popularity of live streams usually grows and drops very rapidly~\cite{dogga2019edge}. Particularly, it often encounters the ``thundering herd'' problem~\cite{federico2015hood, 2016Facebook}: a large number of users, sometimes in the scale of millions, may start to watch the same live video simultaneously when some popular events or online celebrities start a live broadcast. Second, live video delivery nowadays has stringent requirements on latency owing to the new breed of live video services that support interactive live video streaming. These services allow the broadcasters to interact with their stream viewers in real-time during the streaming process. In order to support the high interactivity, it requires low-latency end-to-end delivery while maintaining the Quality of Experience (QoE) for live viewers~\cite{wang2016anatomy, yi2019acm, pang2018optimizing}.

Typical thundering herd problem in live video can overload the system, causing lags and disconnections from the server. One efficient way to solve the thundering herd problem while maintaining low latency in delivery of live videos is to utilize edge caches. For example, Facebook uses edge PoPs distributed around the globe for the delivery of their live traffic~\cite{federico2015hood}. Delivering contents via edge devices (e.g., edge servers co-located with mobile base stations) makes contents much closer to the end users and alleviates the traffic burden of backbone networks to the cloud. 

\begin{figure}[!t]
\begin{center}
\includegraphics[width=1.0\columnwidth]{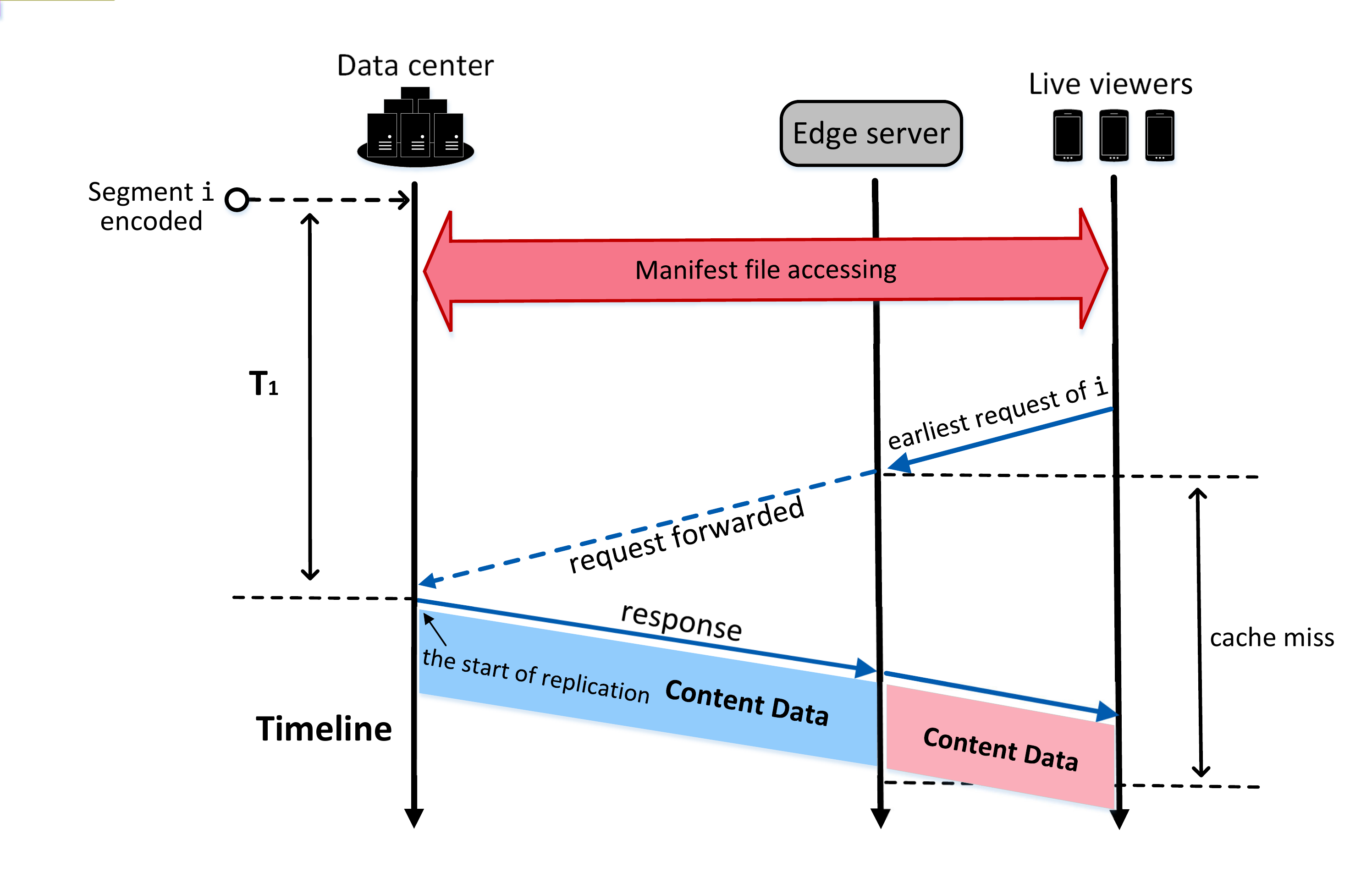}
\caption{Client-driven content caching (replication) for live videos.}\label{fig:motivation_gap}
\end{center}
\vspace{-0.28in}
\end{figure}

Nevertheless, when applying edge-assisted live video delivery, a new problem of cache miss arises: when a large number of end users request for a newly generated video segment at the same time, this segment may not has enough time to be cached in the edge caches due to the real-time property of live streaming~\cite{rainer2016investigating, ge2018qoe}. As shown in Fig.~\ref{fig:motivation_gap}, the edge server would return a cache miss for the first group of requests that arrive the edge before the segment is fully cached. These cache-missed requests would pass the edge cache and go all the way to the origin cache or server. As a result, it would lead to deteriorated QoE to the live viewers (e.g., increased startup latency and playback stall rates). According to the statistics of Facebook~\cite{federico2015hood}, around 1.8\% of their Facebook Live requests encountered cache miss at the edge layer, and cause failures at the origin server level. Note that this is still a significant number considering the magnitude of the number of live viewers. To make matters worse, high revolution videos (e.g., virtual reality (VR) streams) which need more time to be replicated to the edge would create a even higher cache miss rate. 

The above caching problem only exists for live video streaming as people typically watch regular videos at different times. Therefore, there are sufficient time for the regular video chunks to be cached with few fast-entered content requests. State-of-the-art researches solve the above problem by holding back the availability of some newly encoded live segments from the playback clients so that the client requests could arrive at the edge after the caching process is finished~\cite{federico2015hood, ge2018qoe}. This strategy while solves the cache miss problem, would however pose extra latency to the live streams which sacrifices the ``liveness'' of delivered videos.

The root cause of the cache miss problem is mainly because the current client-driven caching strategy was not designed for live videos in the first place. Since caching process in the current content delivery networks (CDNs) is normally triggered by the client requests, so the video segments caching (replication) will only commence when the cloud responses to the first request for a live video segment. While this strategy makes sense when delivering regular content, it slows down the caching process in the context of live videos: there exists a time \textit{gap} (shown as $T_1$ in Fig.~\ref{fig:motivation_gap}) between the time when a segment is generated from the cloud and when the caching process really started. This gap mainly consists of two parts: i) the time that the availability information of the newly encoded video segments is obtained to the playback clients, and ii) the time it takes for the clients to send their first segment request. However, in the current pull-based CDN architecture, both of these two parts of time are difficult to narrow down (refer to~\cref{sec:motivation_relatedWork} for more details). This motivates us to rethink the caching design of live video delivery. Can the cloud CDN server adopts a video push model to proactively replicate the newly encoded video segments into the appropriate edge servers in real-time?

Although desirable, it is challenging to achieve such goal due to the massive video requests and edge servers, QoE guarantee, and high real-time requirement. First, in order to adopt the proactive caching strategy, we must solve the allocation problem between edge servers and live viewers (i.e., assign the viewers to the proper edge server). This is because: i) the video segments that need to be replicated in an edge server is determined by the live viewers served by this edge server, and ii) as the bandwidth capacity of edge servers is quite limited (much smaller than CDN servers), the workloads of many edge servers could be easily overwhelmed while the others are under-utilized. Conventional load balance solutions~\cite{xu2013joint, narayana2012coordinate} which assume that content replicas are stored over all CDN servers would be unrealistic in our context considering the massive number of edge servers. Second, since the service capability of each individual edge server is limited, newly encoded video segments have to be replicated to massive edge servers to alleviate the spiky live video traffic. For each live video segment that being encoded in real-time from the cloud, we need to make a fast decision on the appropriate edge servers to cache the segment.

In this paper, we propose a \underline{p}roactive \underline{l}ive \underline{v}ideo \underline{e}dge \underline{r}eplication scheme (\textit{PLVER}) to resolve the cache miss problem in live video delivery. \textit{PLVER} first conducts a \textit{one-to-multiple} stable allocation between edge clusters and user groups which balances the load of live requests over edge servers such that each user group could be assigned to its most preferred edge cluster that it could be matched. Then based on the allocation result, \textit{PLVER} proposes an efficient and proactive live video edge replication (push) algorithm to speed up the edge replication process by using real-time statistical viewership of the user groups allocated to the cluster. 

In summary, this paper makes the following contributions:
\begin{itemize}
    \item \textit{PLVER} implements a stable \textit{one-to-multiple} allocation between edge clusters and user groups (i.e., one user group is served by one edge cluster but one edge cluster can serve multiple user groups), under the constraint that the QoE of end users is guaranteed by their assigned edge clusters.
    \item Aiming at speeding up the edge replication process, \textit{PLVER} identifies the unique traffic demand of live videos and develop a proactive video replication algorithm to provide fast and fine-grained replication schedule periodically. To the best of our knowledge, this is the first research work to provide proactive video replication algorithms (with details disclosed to the public) tailored for edge-assisted live video delivery.\nop{(Refer to~\cref{sec:relatedlive} for a solution of the same kind from Facebook).} 
    \nop{\item We propose a novel algorithm \textit{PLVER} to solve the video replication problem (i.e., determining the video segments to be cached on edge servers for a certain time). \textit{PLVER} adopts a heuristic approach to optimize the video replication such that live video traffic served by the edge servers could be maximized, subject to a given constraint on total replication costs.}
    
    \item We perform comprehensive experiments to evaluate the performance of \textit{PLVER}. A trace-driven allocations between $641$ edge clusters and $1253$ user groups are conducted, which cover $64$ ISP providers and $470$ cities. Then based on the allocation results, we further evaluate the performance of the video replication algorithm using traces of $0.3$ million Twitch viewers and more than $300$ Twitch channels. Performance results demonstrate the superiority of \textit{PLVER}.
\end{itemize}

\section{Motivation and Related Work}\label{sec:motivation_relatedWork}

\begin{figure*}[!t]
\begin{center}
\includegraphics[width=0.75\textwidth]{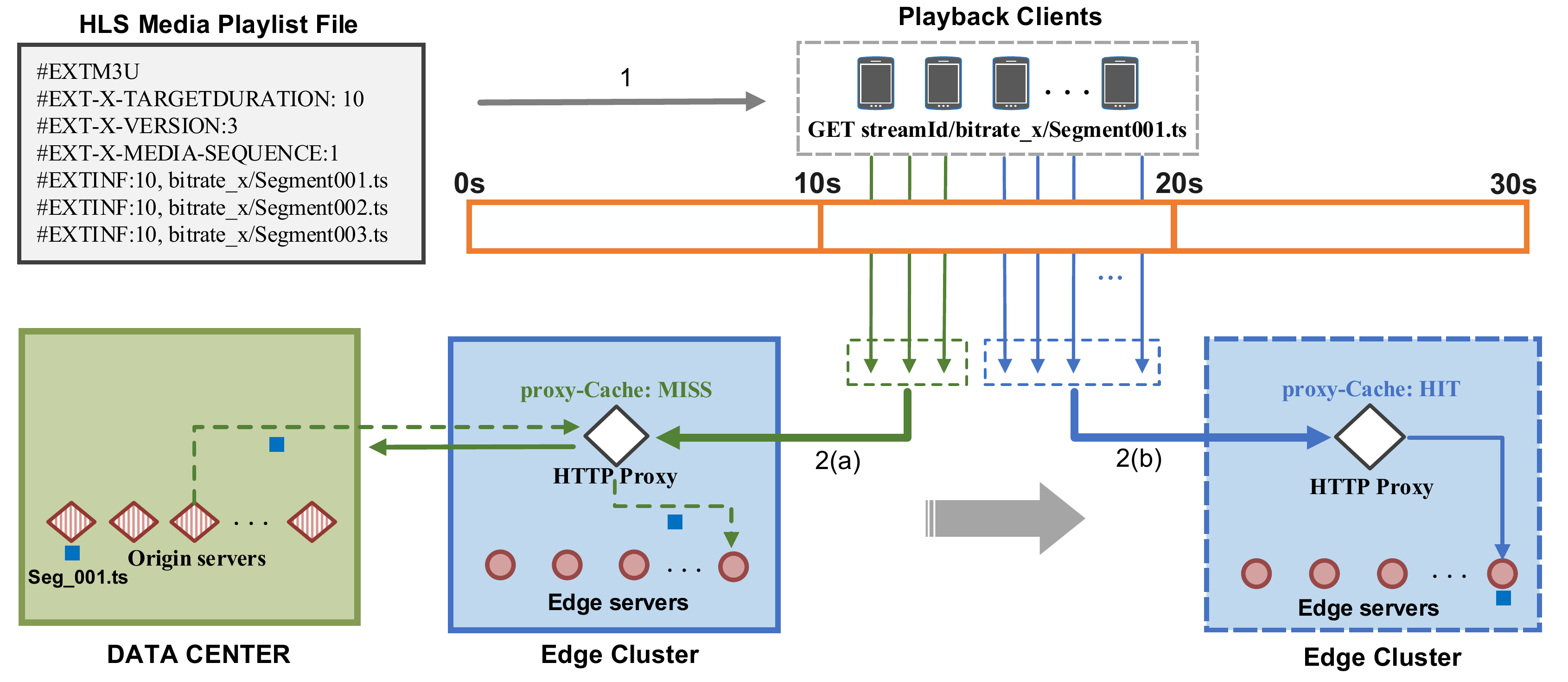}
\caption{Illustration of cache miss problem for edge-assisted live video delivery.}\label{fig:liveVideo_architec}
\end{center}
\vspace{-0.05in}
\end{figure*}

\subsection{Live Video Delivery Background}
A live stream is usually encoded into multiple pre-determined bitrates once it is generated and uploaded by the broadcasters. For each bitrate of a stream, it is further split into a sequence of small video segments with the same playback length, so that it can be fetched sequentially by playback clients (e.g., via \texttt{HTTP} \texttt{GET}), using a suitable bitrate matching their network conditions~\cite{sodagar2011mpeg}.

In the HTTP-based live video delivery, every time when a client joins a live channel, she first request and accesses stream's playlist file (generated by the origin streaming server). This manifest contains the information of current available segments (i.e., segments that have been encoded in the cloud) and bitrates in the stream. Based on the information from the manifest, the clients send the HTTP requests to their local edge server. Afterwards, the playback client fetches the live video segments in sequence and periodically accesses the newest playlist file to check if any new segments have been produced. When live videos delivered over edge servers (as shown in Fig.~\ref{fig:motivation_gap}), these video segments will then be replicated (cached) to the edge caches when the edge HTTP proxy receives the response (segment) from the cloud. 

\subsection{Observation and Motivation}
To better explain the cache miss problem, we use Apple HLS (HTTP Live Streaming) protocol~\cite{rfc8216roger} as an example to illustrate the live video delivery process. As shown in Fig.~\ref{fig:liveVideo_architec}, start from a certain time after $10^{th}$ second of a live stream, the first three video segment were generated from the cloud. By first accessing the playlist file, numerous clients (with geographical proximity) realize the segment update and begin to request segment $001.ts$ via \texttt{HTTP} \texttt{GET} during $10$ to $20$ seconds. These requests would first be handled by one of the HTTP proxies in an edge cluster, which checks if the requested segment is already in an edge cache. If the segment is in the edge cache, then it could be readily fetched from there (step $2(b)$). If not, the proxy will issue a HTTP request to the origin server in the cloud (step $2(a)$). (Note that there exist another layer of cache as well as proxy and encoding servers inside the data center. As our system design does not change the current structure within the data center, these components are dismissed in Fig.~\ref{fig:liveVideo_architec}.)

As we can easily find that an earlier fraction of requests (shown as step 2(a) in Fig.~\ref{fig:liveVideo_architec}) before the segment is fully cached in the edge would miss the edge cache. The current client-driven caching architecture creates a time gap before the caching process is started, which is critical for the live video delivery with real-time requirement. The playback clients request and access the playlist file occupies the first part of time of the gap, which is inevitable in the client-driven content caching since the clients have to know the segment information (i.e., the URI) before sending the requests. Once a video segment availability information is obtained by the playback clients, it takes another period of time before the first request for the segment is sent out by the clients. This part of time exists because the current live streaming protocols (e.g., HLS or MPEG-DASH) would generally start a live streaming with an relatively ``older'' video segment instead of the newest one to avoid playback stalls~\cite{rfc8216roger}. As shown in Fig.~\ref{fig:liveVideo_architec}, the playback clients would start the live streaming by first requesting segment $001.ts$ rather than segment $003.ts$, which makes the replication of segment $003.ts$ further postponed in the client-driven caching architecture.

\begin{figure*}[!htb]
\begin{center}
\includegraphics[width=0.92\textwidth]{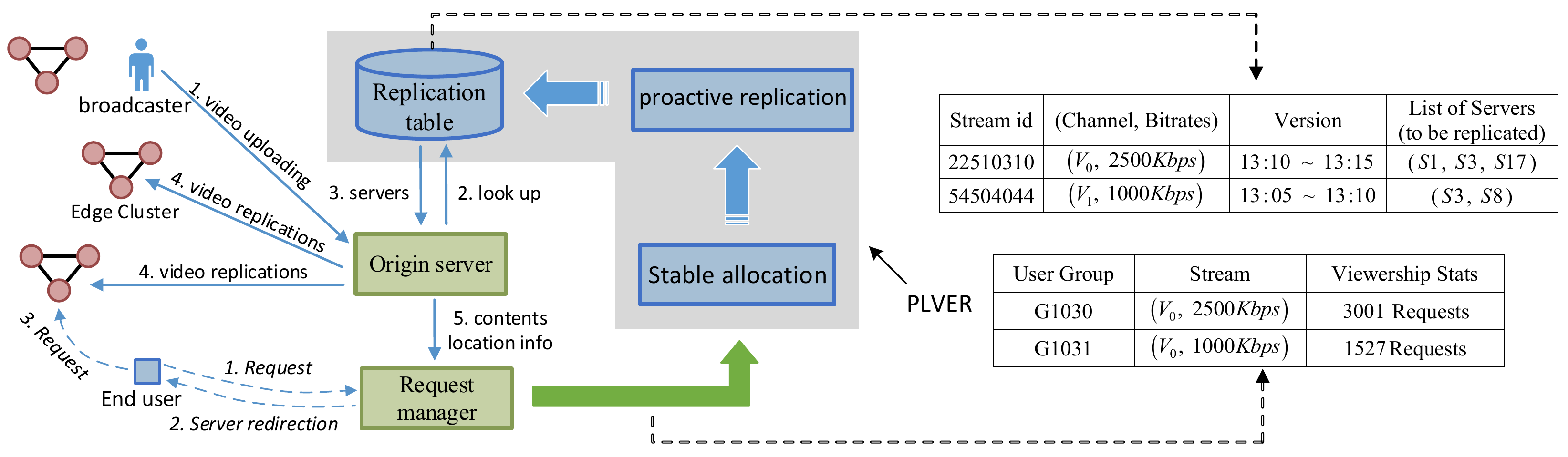}
\caption{System architecture: solid lines denote the procedure for video replication; dash lines denote the procedure that a user accesses live video.}\label{fig:sys_architec}
\end{center}
\end{figure*}

\subsection{Improving the QoE of Live Streaming} \label{sec:relatedlive}
In order to solve the cache miss problem as well as to improve the QoE of $4$K live videos, Ge et al.~\cite{ge2018qoe} proposed ETHLE, which ``holds back'' the availability of some newly encoded video segments from the playback clients so that the playback clients could send their live requests to a certain segment after it was cached in the edge server. While this work has shown considerable QoE improvement, it may pose extra undesirable latency to the live streams. 

In the industry, Facebook proposed two alternative methods to solve the cache miss problem for delivering live video over edge servers~\cite{federico2015hood}. In the first scheme, their solution uses the similar ``holding back" idea as that in~\cite{ge2018qoe}: the edge proxy returns a cache miss for the first request while holding the rest requests in a queue. Once the segment is stored in the edge cache via the HTTP response of the first request, the requests in the queue can be responded from the edge as cache hits. Similar with the work in~\cite{ge2018qoe}, this design would incur undesirable latency to the live stream. The other scheme adopts a video push model where the server continuously pushes newly generated video segments to the proxies and the playback clients. This is the only reported design that adopts proactive content push for live video over edge servers. Nevertheless, the exact details of their video replication algorithm are unknown.

In~\cite{yan2017livejack}, Yan et al. proposed LiveJack, a network service which allows CDN servers to leverage the ISP edge cloud resources to handle the dynamic live video traffic. Their work mainly focus on the dynamic scheduling of Virtual Media Functions (VMFs) at the edge clouds to accommodate with the dynamic viewer populations. Wang et al. proposed an edge-assisted
crowdcast framework which makes smart decisions on viewer scheduling and video transcoding to accommodate with \textit{personalized} QoE demands~\cite{wang2019intelligent}. Mukerjee et al. in~\cite{mukerjee2015practical} performed end-to-end optimization of live video delivery path, which coordinates the delivery paths for higher average bitrate and lower delivery cost. This work, however, mainly focuses on optimizing the routing of live video delivery. In~\cite{zhang2018proactive}, Zhang et al. provided a video push mechanism to lower the bandwidth consumption of CDN by proactively sending the videos to competent seeds in a hybrid CDN-P2P VoD system. This work uses proactive video push, but it does not target at live videos. The optimization for regular, non-live videos delivery was also investigated in~\cite{joseph2014nova, kim2016quality} and~\cite{lu2018optimizing}.

\subsection{Generic Video Replication Techniques}
Different content replication strategies were developed in~\cite{hu2016joint, ma2017joint, zhou2015video} and~\cite{al2018edgecache}. In~\cite{hu2016joint}, Hu et al. considered both video replication and request routing for social videos. Their algorithm focuses on social videos and the watching interests of different communities. In~\cite{ma2017joint}, Ma et al. considered the video replication strategies in edge servers. They proposed a content replication algorithm to jointly minimize the accumulated user latency and the content replication cost. In~\cite{zhou2015video}, Zhou et al. investigated how the popularity of video changes over time and then designed the video replication strategies with the video popularity dynamics derived. Different from ours, the above works mainly focus on the video-on-demand (VoD) services. \nop{Hung et al.~\cite{hung2018combinatorial} also considered the caching allocation problem at edge servers for live videos. Their system model, however, largely differs from ours. They assumed that the bandwidth from the wireless base station (i.e., the edge) to the users is not a constraint due to increased wireless bandwidth in the 5G system but the backhaul link from the base station to the stream services at the cloud may be congested.}

\section{System Overview} \label{sec:back}
Our system design of \textit{PLVER} is shown in Fig.~\ref{fig:sys_architec}. Once a live viewer sends a \texttt{HTTP} request to the \textit{request manager} of the system, the request manager identifies the key information of the request, including the requested channel, bitrates, and the user group it belongs to, by resolving the \texttt{URL} and the \texttt{IP} address. The above information is used by the request manager to redirect the request to an appropriate edge server. This procedure is denoted with blue-dash lines in Fig.~\ref{fig:sys_architec}. The request manager also generates the viewership information (e.g., the number of viewers of each stream in each user group) and feeds the information to \textit{PLVER} for edge servers selection. 

As shown in Fig.~\ref{fig:sys_architec}, there are three main components \textit{PLVER}: i) \textit{stable allocation} module assigns the global user groups to their desired edge server cluster and balance the load of live traffic, ii) the \textit{proactive replication algorithm} periodically computes the edge replication schedule within each edge cluster in the near future (e.g., next $5$ minutes), based on the viewership information from the request manager, and iii) \textit{replication table} which contains the directly available information of replication servers for each live video segment.

When the new live video segments of a stream are encoded and generated, the system first checks the most up-to-date replication schedule from the replication table by identifying the key information of the segment. It then proactively replicates these video segments into the guided edge servers despite these videos are currently not requested by the users. In this way, replication schedule can be obtained easily and fast by using the stream id and version number of the target video segment as the key for searching. Note that the process of replicating the video segments into edge servers and delivering the video contents from edge servers to the end users are conducted concurrently, since the video segments are generated from the broadcasters sequentially. \nop{The above procedure is denoted with solid lines in Fig.~\ref{fig:sys_architec}.}

The core component of the system is \textit{PLVER}, denoted in the grey box in Fig.~\ref{fig:sys_architec}. Its main goal is to provide replication schedule that can be readily used for live video replication over edge servers. To be more specific, it considers the traffic demand from different areas as well as resource capacity of edge servers so as to provide replication schedule that maximizes the traffic served by the edges. Note that tracking each viewer's requests and directing the requests to edge servers or the origin server belong to real-time request redirection. It happens after the replication schedule is generated and needs to consider the dynamic content availability in edge servers, which is beyond the scope of this paper. Nevertheless, it will be utilized for performance evaluation of our algorithm in the evaluation part (\cref{sec:evaluation}) of this paper.

In the following, we formally model the problem that needs to be solved by \textit{PLVER}, and then present the two main components of our solution, namely \textit{stable one-to-multiple allocation} and \textit{proactive replication algorithm}, in~\cref{sec:stable_allocation} and~\cref{sec:per}, respectively. 

\section{Stable One-to-multiple Allocation} \label{sec:stable_allocation}

\subsection{The Allocation Problem}
\textit{Instead of making the request routing decisions individually for each client, we conducted the servers allocation at the granularity of user groups}. The users in the same group generally have the same network features (\textit{e.g., subnet, ISP, location}) and thus are likely to experience similar QoE when dispatched to the same server~\cite{niereducing,sun2016cs2p}. Similarly with conventional content delivery problem, it is generally necessary to first consider the load balance problem between edge server clusters (consisting of a number of edge servers with the same network features) and the user groups. 

We consider a target network of a number of edge server clusters and user groups. Each user group $i$ originates an associate live traffic demand $d_i$, and each edge cluster $j$ has a capacity $C_j$ to serve the demands. In order to satisfy the QoE of users, for each user group, it has a list of candidate edge clusters in descending order of preference. A higher preference indicates those clusters that can provide better predicted performance for the viewers in the group (e.g., lower latency and packet loss). Likewise, each edge cluster $j$ also has preferences regarding which map units it would like to serve~\cite{maggs2015algorithmic}.

An allocation of edge clusters to user groups is said to be a \textit{stable marriage} if there is no pair of participants (i.e., edge clusters and user groups) that both would be individually better off than they are with the element to which they are currently matched~\cite{gusfield1989stable}. By conducting stable allocation, each user group is assigned to its most preferred server cluster to which it could be assigned in any stable marriage. In other words, stable allocation implies the most desirable matching between user groups and server clusters. The goal of our allocation problem is to assign the user groups to the edge clusters, such that the capacity constraints are met and the bidirectional preferences are accounted for. 

\subsection{Stable Allocation Implementation Challenges}

However, in the context of live video delivered over edge servers, the stable allocation has practical implementation challenges listed below.
\subsubsection{Expensive many-to-many assignment}
Conventional allocation used by CDN vendors normally generates a many-to-many assignment, i.e., the traffic demand of each user group could be served by multiple edge clusters. Many-to-many assignment makes sense when there are only a small number of server clusters globally. In our context, however, the number of edge clusters is much more than that of conventional CDN clusters, thus a many-to-many assignment becomes unnecessarily expensive.
\subsubsection{Partial preference lists}
Considering the large number of edge clusters and user groups, it is unnecessary to measure and rank the preference of every edge cluster for each user group. Therefore, there is a partial preference lists of edge clusters that are likely to provide the best performance for a given user group. Similarly, the edge clusters also only need to express their preferences for the top user groups that are likely candidates for assignment.
\subsubsection{Integral demands and capacities}
The canonical implementation of stable marriage problem considers unit value demands and capacity, while in our case the demands of user groups as well as the capacities of server clusters could be arbitrary positive integers. 

\setlength{\textfloatsep}{6pt}
\begin{algorithm}[!htb]\small
\caption{\textit{ISOA}}\label{algo:ext_galeShapley}
\KwIn {Preference list by user group and edge cluster: $uP$, $cP$; $C_j$: resource capacity of an arbitrary edge cluster $j$; $D()$: traffic demand of given user groups.}
\KwOut{$G_j$: List of allocated user groups to edge cluster $j$.}
Initialize all user groups as free; 
$G_j = \{j: \text{[],  for $j$ in } E\}$\;
\ForEach {$i\in$ free user groups}{
    $j \leftarrow$ head of $uP_i$\; 
    Insert $i$ into $G_j$ (rely on $cP_j$)\;
    \If {$D(G_{j}) <= C_j$}{
        continue\;
    }
    $start$ $\leftarrow$ bSearch($G_j, C_j, i$); $k \leftarrow start + 1$\;
    \While {$k \leq length(G_j)$}{
        \If{$D(G_j^{0\thicksim k}) > C_j$}{
            remove $G_{j}^k$ from $G_j$\;
            \If {$G_{j}^k == i$}{
                remove $j$ from $uP_i$; goto line $4$\; 
                
            }
            \Else{
                label $G_{j}^k$ as free user groups\;
            }
        }
        $k \leftarrow k + 1$\;
    }
}
\Return $G$;
\vspace{-0.05in}
\end{algorithm}

\begin{algorithm}[!htb]\small
\caption{bSearch($G_j, C_j, i$)}\label{algo:bSearch}
$left \leftarrow$ position of $i$ in $G_j$; $right \leftarrow$ length of $G_j$\;
$mid = \frac{left + right}{2}$\;
\While{True}{
    \If{$D(G_j^{0\thicksim mid}) \leq C_j$}{
        \If{$D(mid + 1) > C_j$}{
            \Return $mid$\;
        }
        \Else{
            $mid = \frac{mid + right}{2}$\;
        }
    }
    \Else{
        $mid = \frac{left + mid}{2}$\;
    }
}
\end{algorithm}

\subsection{Solution Methodology}

To address the above challenges, we propose a new allocation algorithm: \textit{Integral Stable One-to-multiple Allocation (\textit{ISOA})}, which extends the Gale-Shapley algorithm used for solving the canonical stable allocation problem. \textit{ISOA} works in rounds, where in each round, each free user group (all user groups are free initially) proposes to its most preferred edge cluster, and the edge cluster could (provisionally) accept the proposal. Let $G_i$ denote the list of user groups assigned to edge cluster $i$. In the case that capacity of edge cluster is violated, we perform a binary search on $G_i$ to identify the user groups that need to be evicted.

Algorithm~\ref{algo:ext_galeShapley} shows the details of \textit{ISOA}, where $uP$ and $cP$ are the preferred list of edge clusters (by user groups) and the preferred list of user groups (by edge clusters), respectively. $C_j$ denotes the service capacity of edge cluster $j$ (Note that in practice, $C_j$ could be a \textit{resource tree} instead of a single value~\cite{maggs2015algorithmic}). To find a stable one-to-multiple allocation, we first set all user groups as free and set the initial user groups to each edge cluster as empty (line 1). Then, we pick up a free user group $i$ in each round and get its most preferred edge cluster $j$ (line 2-3). Based on the preference of edge cluster, we insert $i$ into the temporarily-assigned user group list of edge cluster $j$. After adding a new user group to $G_j$, the current traffic demand needed by $G_j$ may or may not violate the capacity $C_j$. If $C_j$ is not violated, we go back to propose another free user group for proposing (line 5-6).

\begin{figure}[!t]
\vspace{-0.1in}
\hspace{-0.17in}
\begin{center}
\includegraphics[width=0.8\columnwidth]{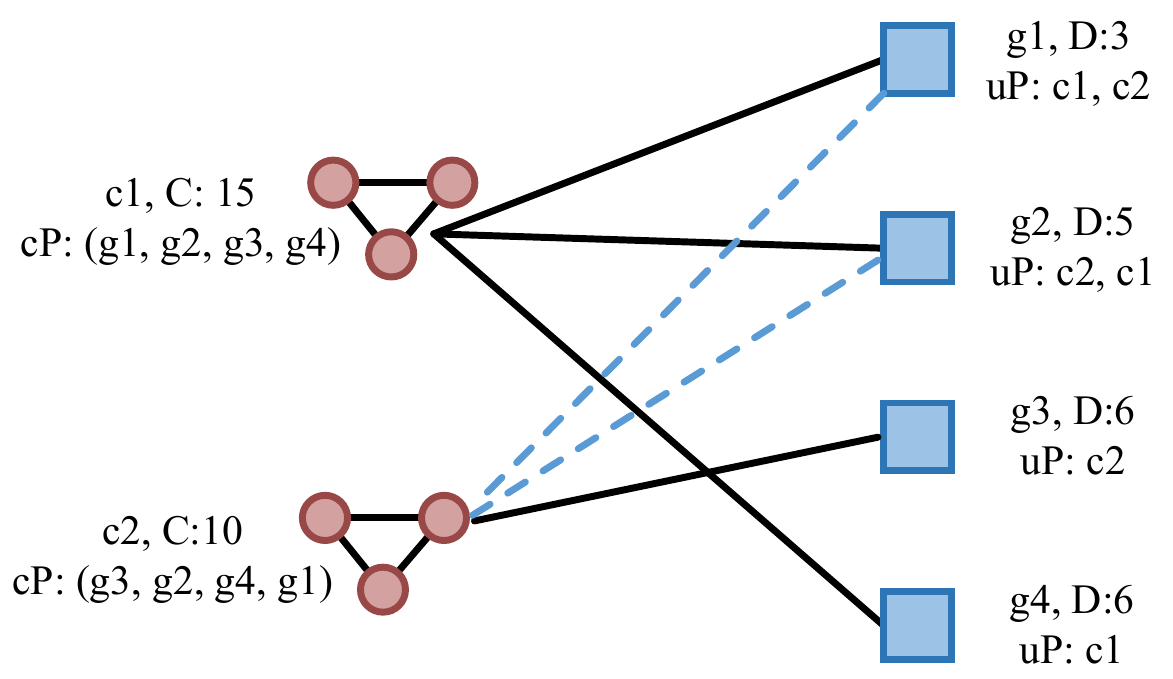}
\vspace{-0.08in}
\caption{An example of the stable one-to-multiple allocation containing four user groups and two edge clusters, where the service capacity of each edge cluster is denoted by 'c' and the traffic demand of each user group is denoted by 'D'}\label{fig:ispa_example}
\end{center}
\end{figure}

In case of $C_j$ is violated, we conduct a binary search (refer to Algorithm~\ref{algo:bSearch}) to find out the first user group ($start + 1$) in $G_j$ which causes the capacity violation. We further go through all user groups from $G_j^{start +1}$ to the end of $G_j$: if adding a user group ($G_j^k$) would cause the capacity violation, then we remove this user group from $G_j$ (line 8-10). If the removed user group is $i$ itself, it suggests that $i$ cannot get its most preferred edge cluster. In that case, $i$ will go back to propose to its second preferred edge cluster (line 11-12). Otherwise, the evicted $G_j^k$ will be labelled as a free user group, waiting for a second-chance proposal.

As a simple example, Fig.~\ref{fig:ispa_example} shows the meaning of the stable one-to-multiple allocation, where we have two edge clusters $c1$ and $c2$, with service capacity of $15$ and $10$, respectively. There are $4$ user groups ($g1$, $g2$, $g3$ and $g4$), which generate traffic demands of $3$, $5$, $6$ and $6$, respectively. The preferred edge cluster list by each user group as well as the preferred user group list by edge cluster are shown in this figure with $uP$ and $cP$, respectively. We need to match each user group to their most preferred edge cluster that it could be assigned to.

Running \textit{ISOA} with the simple example in Fig.~\ref{fig:ispa_example}, user group $g1$, $g3$, $g4$ can propose and be matched to their most preferred edge cluster ($c1$, $c2$ and $c1$, respectively). Group $g2$, however, will trigger the capacity violation of $c2$, thus can only be matched to its second preferred cluster $c1$. The matching results are marked with the solid lines in Fig.~\ref{fig:ispa_example}.  \nop{ As a result, \textit{ISOA} will return $G_{c1} = \{g1, g2, g4\}$ and $G_{c2} = \{g3\}$  as the allocation result, which is marked with solid lines in Fig.~\ref{fig:ispa_example}.}

\begin{table}
    \centering
    	\caption{Summary of main notations in~\cref{sec:per}}\label{tbl:notations}
    	\vspace{-0.051in}
	\begin{tabular}{ | m{1cm}<{\centering} | m{6.5cm}|}
		\hline
		\emph{Notation}        & \emph{Description}  \\ 
		\hline
		\hline
		$F$             & Target edge cluster within which the replication problem to be solved\\
    	$E$             & Set of all edge servers over $F$\\
    	$U$             & Online viewers from the user groups assigned to $F$ \\
    	$T$             & Target time window in the near future\\
    	\hline
    	\vspace{0.03in}
    	$A_j$           & Set of viewers that are served by edge server $j$\\
    	$B_j$           & Bandwidth capacity of edge server $j$\\
    	$a_i$           & The edge server that serve viewer $i$ during $T$\\
    	$b_i$           & Bandwidth consumed by viewer $i$\\
	    $s_i$           & The live stream that viewer $i$ is watching\\
	    $c_j$           & Cache capacity of edge server $j$.\\
    	\nop{$C_j$           & Cache capacity of edge server $j$\\
    	$G_i$           & The user group that viewer $i$ belongs to\\}
    	\hline
    	\vspace{0.03in}
    	$\mathcal{D}_{i}^{T}$ & The video segments to be generated by $s_{i}$ in $T$\\
    	$L_{A_j}$         & Non-redundant set of streams accessed by viewers in $A_j$\\
        $S(\cdot)$           & Size function that calculates the total data volume in a set of video segments\\
        $\mathcal{V}_j^T$   & Live video segments that should be replicated into edge server $j$ during $T$\\
        $L(v_{i}^t)$    & The replication schedule: list of edge servers that video segments $v_i^t$ should be replicated into\\
    	\hline
	\end{tabular}
\end{table}

\section{Proactive Replication over the Edge}\label{sec:per}

\subsection{Notations and Assumptions}
Once the allocation problem is solved, we only need to consider the replication problem within each single edge cluster and its assigned user groups (Note that the QoE of the assigned user groups is guaranteed with stable allocation). We next formulate the single cluster replication problem by considering a given edge cluster $F$ and the user groups assigned to $F$. The main notations used in our problem formulation are listed in Table~\ref{tbl:notations}. Without loss of generality, we make the following assumptions:
\begin{itemize}
    \item We consider a target time window $T$ in the near future that we need to generate the video replication schedule. During $T$, a number of live streams are watched by the live viewers $U$ distributed across the user groups allocated to $F$. \nop{Each live channel could be further encoded into different live streams with different bitrates (video quality).}
    \item Each end user is served by one edge server at a time, and clients that cannot be served by the edge servers will be directed to the cloud.
    \item The cache in an edge server can be shared by multiple viewers who are accessing the video, but each viewer consumes their exclusive bandwidth of the edge server. \nop{In other words, the cache capacity of edge server is sharable, while the bandwidth of edge server is not sharable~\cite{he2018its}.}
\end{itemize}  

\subsection{Resource Constraints}

We divide time into a series of short, consecutive time windows, and try to generate video replication schedule for each time window based on the feed of most up-to-date viewership of live videos. The goal of the replication schedule is to maximize the traffic served by the edge servers so as to improve the QoE of end users. 

Since clients generally access the video segments of a live stream sequentially, users' demands to the video segments to be generated in the next short time window can be roughly estimated by the current viewership of this stream. Note that the live viewers might change their video quality (bitrates) during the watching process, while the distributed design and a fine-grained time window allow the system to quickly respond to stream demand change. We use $a_i$ to represent the edge server that serves viewer $i$ in $T$, and use $A_j$ to denote the set of viewers served by server $j$, i.e.,
\begin{equation}\label{eqt:def_aj}
A_j := \{i| a_i = j, \forall i \in U\}.
\end{equation}
Since each viewer only needs to be served by one edge server, we have
\begin{equation}\label{eqt:single_server}
A_{j_1} \cap A_{j_2} = \varnothing, \forall j_1 \neq j_2.
\end{equation}
For an arbitrary edge server $j$, it should have enough bandwidth to serve $A_j$. Thus, the following constraint should be posed:
\begin{equation}\label{eqt:resource_constr1}
\sum_{i\in A_j}b_i \leq B_j, \forall j \in E,
\end{equation}
where $b_i$ is bandwidth consumed by viewer $i$, and $B_j$ is the total bandwidth of edge server $j$.

Let $s_i$ denote an arbitrary live stream of one live channel with a certain bitrate. We denote the video segments to be generated by $s_{i}$ in $T$ as $\mathcal{D}_{i}^{T}$ (i.e., $\mathcal{D}_{i}^T = \{v_{i}^{t_1}, v_{i}^{t_2},\ldots,v_{i}^{t_n}\}$, where $(t_1, t_2,\ldots, t_n)$ are the timestamps of the video segments $(v_{i}, v_{i},\ldots,v_{i})$ generated in $T$, respectively). If we use $L_{A_j}$ to denote the non-redundant set of streams accessed by viewers in $A_j$ (note that $|L_{A_j}| \le |A_j|$ as one stream is normally watched by more than one viewers), the following constraint on cache capacity should be posed:
\begin{equation}\label{eqt:resource_constr2}
\sum_{i\in L_{A_j}}S(\mathcal{D}_{i}^T) \leq c_j, \forall j \in E,
\end{equation}
where $S(\cdot)$ is the function that calculates the total caching size of a given set of video segments, $c_j$ denotes the cache capacity of edge server $j$.

\subsection{Cost of Content Replication}\label{subsec:replicationCost}
While edge servers benefit the live viewers, we may need to generate multiple replicas of single video content over the edge servers. More replicas on the edge servers generally mean more cost of cache resources at the edge as well as extra delivery cost from cloud to the edge servers.

To reach a good balance, we pose the following constraint to limit the overall replication cost: 
\begin{equation}\label{eqt:rep_cost}
\sum_{j \in E}\sum_{i\in L_{A_j}}S(\mathcal{D}_{i}^T) \leq \alpha \cdot \sum_{j \in E} c_j,
\end{equation}
where $\sum_{j \in E}\sum_{i\in L_{A_j}}S(\mathcal{D}_{i}^T)$ is the total size of overall replicas cached in the edge servers, and $\sum_{j \in E}c_j$ represents the total size of videos that could be cached globally. We use $\alpha$, a percentage variable, to bound the total amount of videos that could be replicated, so as to limit the video replication cost.
\nop{
\subsection{QoE Guarantee} 
When serving end users by edge servers, each user group usually has a list of their preferred edge servers. The preferred edge servers of a user group are those edge servers that can provide the required performance for the clients in that group. For instance, the preferred edge servers for a user group in Boston should be the edge servers near Boston rather than those in Tokyo. Therefore, a viewer $i$ in user group $G_i$ should not be served by an edge server not in the preference list of $G_i$, i.e., 
\begin{equation}\label{eqt:qoe}
i \notin A_j \text{, if } G_i \notin M_j, 
\end{equation}
where $M_j$ is the set of user groups which have edge server $j$ in their preference list.

\begin{remark} The online viewers in the same user groups while accessing the same live stream are equivalent to each other in our problem, since they generate the same traffic demand to the same videos and have the same QoE when served by a server. Therefore, viewer $i$ in the main notations could refer to any online viewer in its user group while watching the same stream, instead of a specific viewer. \nop{The design variables ($a_i, A_j$) are used as the method to estimate the traffic demand so as to design good replication schedules accordingly, instead of fixed bindings between edge servers and end users when conduct real-time request dispatch afterwards.}
\end{remark}
}
\subsection{Problem Formulation} 
The problem needs to be solved by the \textit{Brain} can be formulated as: 
\nop{Aiming at maximizing the live video traffic carried by edge servers, we formulate our problem as follows:}
\begin{subequations}\label{optimization}
\begin{align}
& \label{opt_target}\underset{\{a_i, A_j\}}{\text{max}} && \sum_{j\in E}\sum_{i \in A_j}b_i\\
& \text{s.t.} && (\ref{eqt:single_server}), (\ref{eqt:resource_constr1}), (\ref{eqt:resource_constr2}), \textit{ and } (\ref{eqt:rep_cost}) 
\end{align}
\end{subequations}
Solving~(\ref{optimization}), we obtain $a_i$ and $A_j$, with which the video replication schedule could be easily derived. That is, the video segments that should be replicated into edge server $j$ during $T$ are given by the following: 
\begin{equation}
\mathcal{V}_j^T = \sum_{i \in L_{A_j}}\mathcal{D}^T_i.
\end{equation}
Based on $\mathcal{V}_j^T$ of each edge server, we can do a simple reverse transformation to get the video replication schedule, i.e., for an arbitrary video segment from stream $i$ with timestamp $t$ ($v_i^t$), the list of edge servers to which it should be replicated in $T$ is given by:
\begin{equation}\label{eqt:reverse_transformation}
L(v_{i}^t) = \{j| v_{i}^t \in \mathcal{V}_j^T, \forall j \in E, \forall t \in T\}.
\end{equation}
This video replication schedule is then inserted into the \textit{Replication table} in Fig.~\ref{fig:sys_architec}, by identifying the channel and bitrate of stream $i$. Problem~(\ref{optimization}) is an integer linear program with a massive number of design variables. In the rest of this section, we present a two-step heuristic algorithm to solve this problem.

\subsection{Solution Methodology}
Since live video traffic is network intensive~\cite{maggs2015algorithmic}, the non-sharable bandwidth constraint is a harder constraint compared with the sharable cache capacity constraint. Therefore, \textit{PLVER} first considers the replication problem while temporarily ignoring the constraint on the cache capacity (\textbf{Step 1}). It then conducts adjustments by moving workloads from the edge servers where the cache capacity constraint is violated to the edge servers with available cache and bandwidth resources (\textbf{Step 2}).

\begin{figure*}[!t]
	\centering
	\subfigure[Number of viewers in the system over time.]{
		\label{fig:nViewers}
		\includegraphics[width=0.3\textwidth]{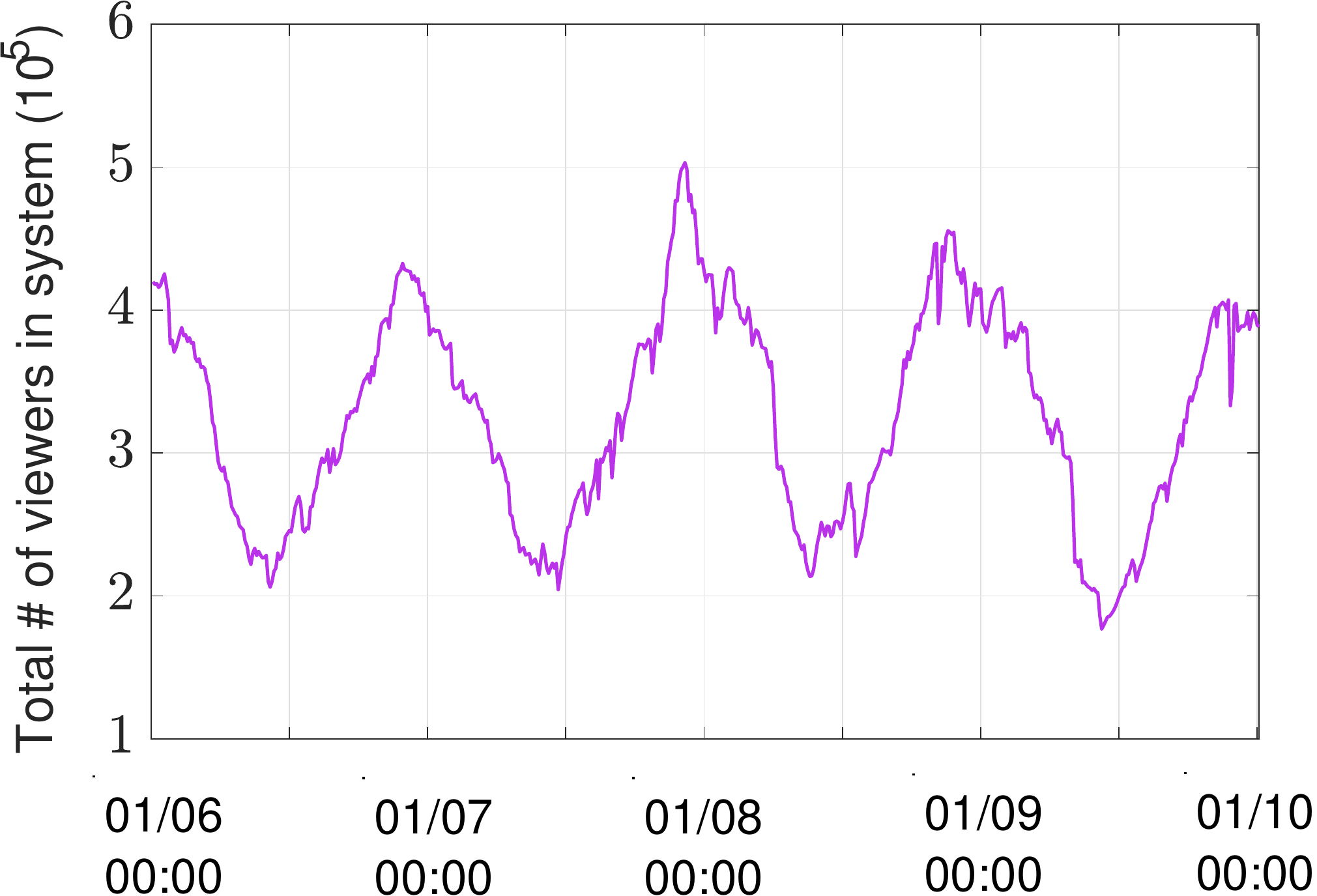}}
	\hspace{0.01in}
	\subfigure[The distribution of number of sessions with different number of viewers.]{
		\label{fig:viewers_dist}
		\includegraphics[width=0.3\textwidth]{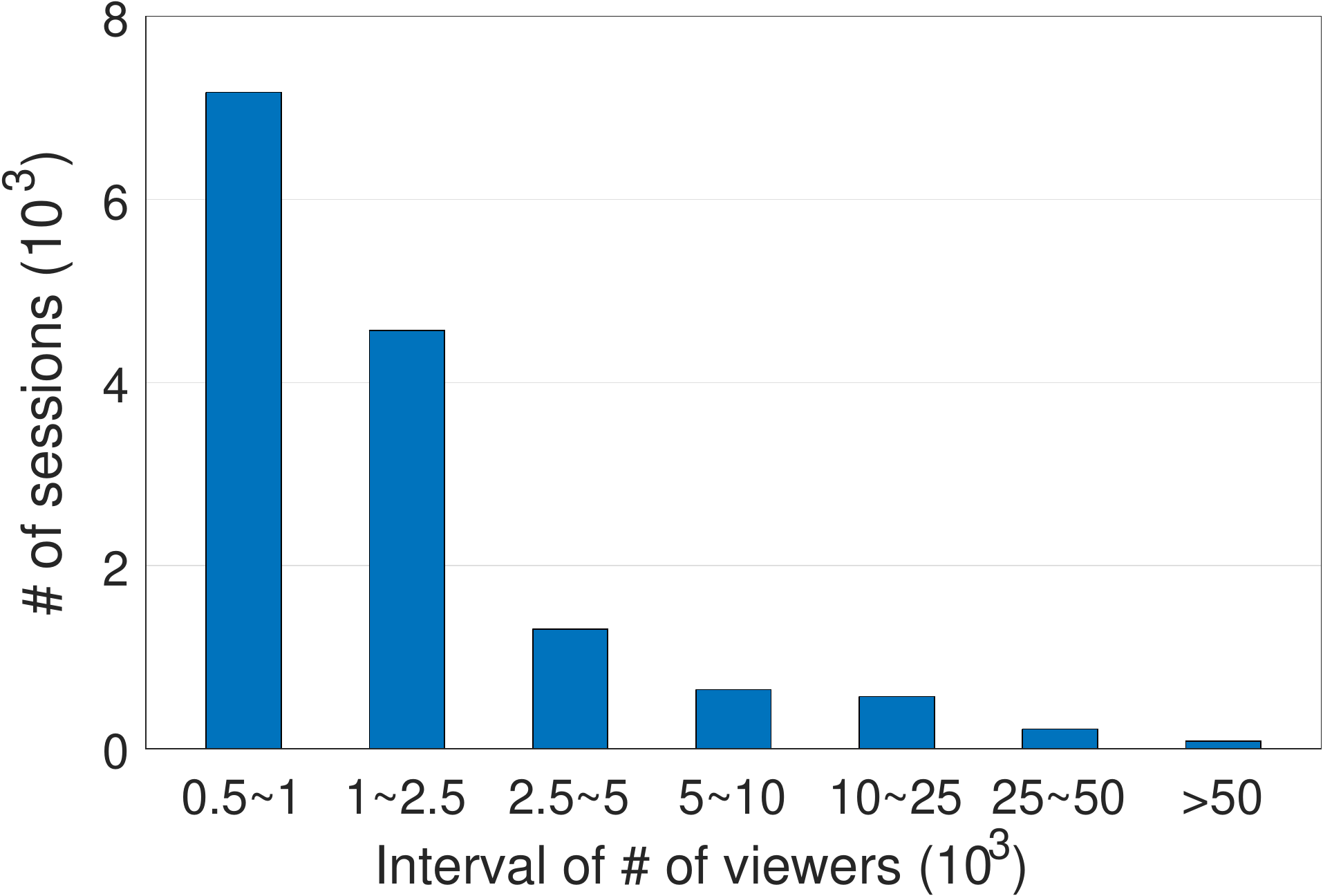}}
	\subfigure[The CDF of the bitrates of live channels.]{
		\label{fig:bitrateCdf}
		\includegraphics[width=0.3\textwidth]{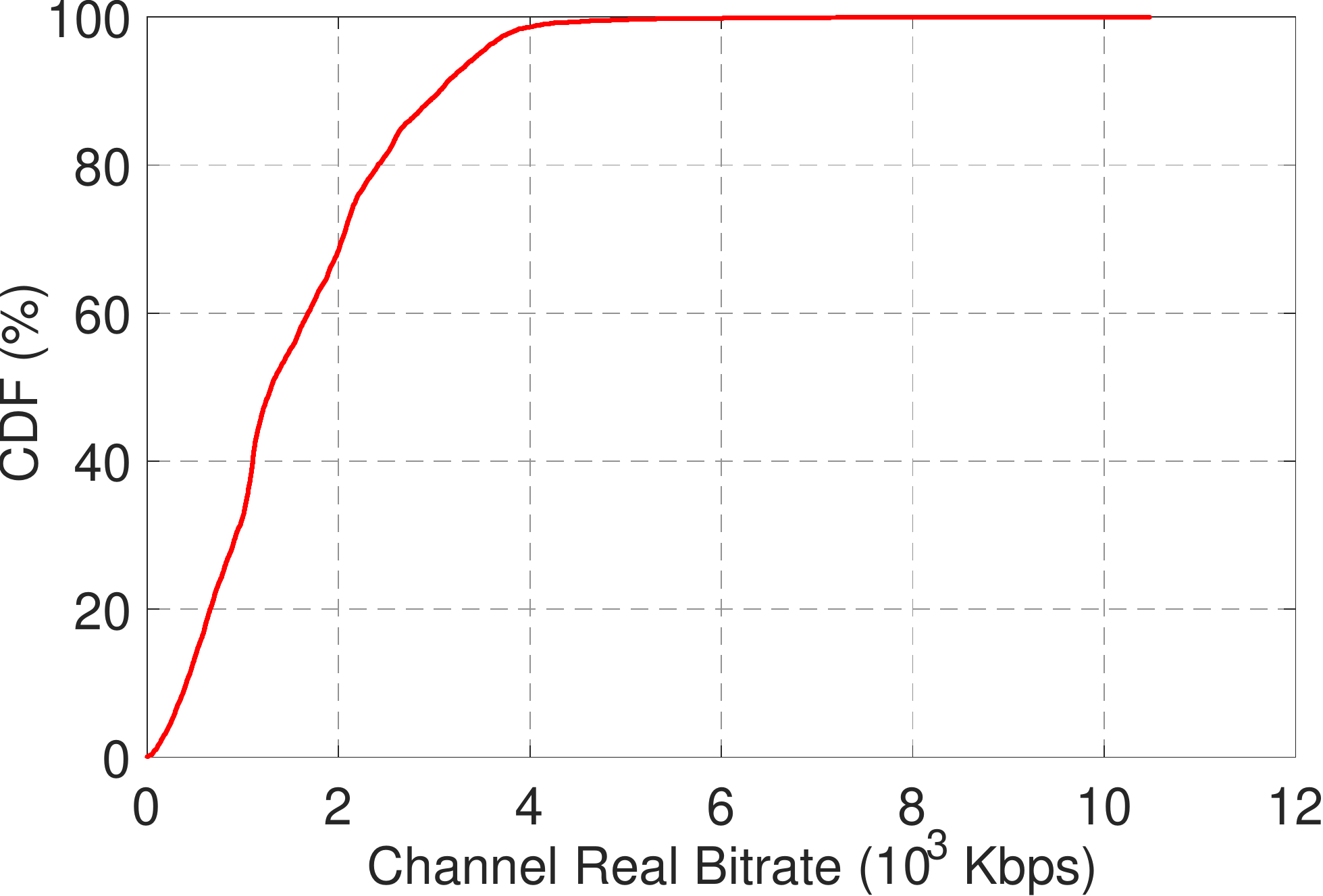}}
	\caption{The statistical information of the experimental dataset.}
	\label{fig:exp_setup}
\end{figure*}

\textbf{Step 1: Greedy Edge Replication for Maximum Traffic: }
By temporarily ignoring the cache capacity constraint, the replication problem could be transformed into the \textit{Multiple Knapsack Problem (MKP)}~\cite{chekuri2005polynomial}. This problem is defined as a pair $\mathcal{(B, S)}$ where $\mathcal{B}$ is a set of $m$ bins and $\mathcal{S}$ is a set of $n$ items. Each bin $j \in \mathcal{B}$ has a capacity $c_j$, and each item $i$ has a weight $w_i$ and a profit $p_i$. The objective is to assign the items to the bins such that the total profit of the assigned items is maximized, and the total weight assigned to each bin does not exceed the corresponding capacity.

If we treat the profit of each viewer $i$ as the bandwidth consumption $b_i$ of that viewer, the \textit{MKP} problem is equivalent to our replication problem with unlimited cache capacities, i.e.,
\begin{subequations}\label{opt-2}
\begin{align}
& \underset{\{x_{ij}\}}{\text{max}} && \sum_{j\in \mathcal{B}}\sum_{i\in \mathcal{S}}b_i x_{ij},\\
& \text{s.t.} && \sum_{i \in \mathcal{S}} b_i x_{ij} \leq B_j,\forall j \in \mathcal{B}\label{opt-2:band_constrait}\\
&&& \sum_{j \in \mathcal{B}} x_{ij} \leq 1, \forall i \in \mathcal{S}\\
&&& x_{ij} \in \{0, 1\}, \forall i \in \mathcal{S}, j \in \mathcal{B},\quad\quad
\end{align}
\end{subequations}
where $\mathcal{B}$ and $\mathcal{S}$ are defined as the set of edge servers in a given edge cluster and the set of viewers assigned by the stable one-to-multiple allocation, respectively, and $x_{ij}$ is defined as follows:
\begin{equation}\label{eqt:s_function}
x_{ij} := \begin{cases}
               1\text{ , if viewer $i$ is served by edge server $j$,}\\
               0\text{ , otherwise}.
            \end{cases}
\end{equation}

The \textit{MKP} problem has been well-researched and has a polynomial time approximation solution (\textit{PTAS})~\cite{chekuri2005polynomial}. Once solving Problem~(\ref{opt-2}), we could further calculate the set of video segments that should be replicated into each edge server based on the value of $x_{ij}$. Let $\mathcal{P}_j$ denote the set of video segments that should be stored in edge server $j$ after Step 1.

\textbf{Step 2: Workload Adjustment:} The solution $\mathcal{P}_j$ can maximize the total amount of traffic served by edge cluster under the assumption of unlimited cache capacities of edge servers. Posing the constraint of limited cache capacity, we need to further adjust the solutions by moving part of the replication workloads from the edge servers whose cache capacity is violated, to the edge servers that have spare cache and bandwidth.

\begin{algorithm}[!t]\small
\caption{\textit{proactive replication algorithm}}\label{algo:scer}
\KwIn{$\mathcal{F}$: given edge cluster; $cc_j$ and $bd_j$: available cache capacity and bandwidth of edge server $j$ ($j \in \mathcal{F}$), respectively; $\mathcal{M}$: set of user groups allocated to $\mathcal{F}$; $s_i$: the live stream accessing by viewer $i$; $b_i$: bandwidth consumption of $s_i$;}
\KwOut{Replication schedule $\mathcal{S}_j$, i.e., for each edge server $j$, the set of video segments should be replicated during the near time window $T$.}

/* Phase 1: Generate initial replication schedule by solving \textit{MKP}. */\\
$x_{ij} \leftarrow$  solvingMKP($bd_j$, $b_i$); $\mathcal{S}_j \leftarrow \varnothing$ , $\forall j \in \mathcal{F}$\\
$a_i$ = $\varnothing$, for all$\text{ viewer } i \in \mathcal{M}$;\\
\ForEach{edgeServer $j \in \mathcal{F}$ }{
    $I = \{i|x_{ij} = 1\}$ //viewers that are served by server $j$;\\
    $L \leftarrow$ the set of unique streams consumed by the viewers in $I$;\\
    sort $L$ according to $\mathcal{R}(s, j)$; (refer to~(\ref{eqt:reward_func})) \label{reward}\\
    \ForEach{stream $s \in L$}{
        \If{$\mathcal{D}_{s}^T.size() > cc_j$}{
            continue;\\
        }
        $\mathcal{S}_j$.append($\mathcal{D}_{s}^T$);
        $cc_j \leftarrow cc_j - \mathcal{D}_{s}^T.size()$;\\
        $a_i = j, ($\textit{for all} $i \in I \textit{ and } s_i = s) $;\textit{ update} $bd_j$;\\
    }
}
/* Phase 2: Redirecting viewers. */\\
\ForEach{viewer $i$ with $a_i= \varnothing$}{
    \If{exists server $j \in \mathcal{F}$ with $bd_j \geq b_i $ \&\& $s_i \in \mathcal{S}_j$}{
        $a_i \leftarrow j$; $bd_j \leftarrow (bd_j - b_i$);\\
    }
}
/* Phase 3: Offloading replication tasks. */\\
\While{exits edge server $j$ with available resources}{
    \If{$a_i \neq \varnothing$ \textit{for all} $i \in \mathcal{M} $ }{
        break;\\
    }
    pick the stream $s$ with max $\mathcal{R}(s,j)$ and $cc_j \geq \mathcal{D}_{s}^T.size()$;\\
    \If{$s \neq \varnothing$}{
        $\mathcal{S}_j$.append($\mathcal{D}_{s}^T$);
        $cc_j \leftarrow cc_j - \mathcal{D}_{s}.size()$;\\
        $bd_j \leftarrow bd_j - \mathcal{R}(s,j)$; $\varphi = \mathcal{R}(s,j) / b$;\\
        \nop{$update$ $a_i$ ($\forall s_i = s$);}
        \textit{update $\varphi$ (number) unassigned viewers with $a_i = j$, $\forall i$ with $s_i = s$};
    }
}
\Return $\mathcal{S}_j$;\\
\vspace{-0.03in}
\end{algorithm}

The whole proactive replication algorithm is shown in Algorithm~\ref{algo:scer}, including three phases. Phase 1 represents Step 1, and phases 2 and 3 represent Step 2 introduced above. Once phase 1 is finished, there might be some viewers whose demand cannot be satisfied (i.e., $a_i = \varnothing$) if we pose cache capacity constraint. For each of these unassigned viewers, in phase 2 we try to redirect it to an edge server with available bandwidth capacity and has the required video segments cached already. There are no directly available edge servers that could be used to serve the rest of the unassigned viewers after phase 2. Thus, in phase 3, the algorithm offloads the incomplete video caching tasks to the edge servers with residual resources. The algorithm returns when all traffic demands are completely satisfied or all edge servers in the given edge cluster are fully loaded. 

The time complexity of our algorithm is ${O}(n+m)$ ($n$ and $m$ are the number of viewers in $\mathcal{M}$ and the number of edge servers in $F$, respectively), without considering the first step of solving the \textit{MKP} problem. Since there are different approximation scheme for solving \textit{MKP} in polynomial time and \textit{PLVER} is a decentralized algorithm with viewers and edge servers from a single edge cluster (i.e., $n$ and $m$ in a small magnitude), \textit{PLVER} could be solved easily.

\begin{remark}\label{remark2} Note that PLVER does not need to track the real-time information of each viewer (e.g., stream being watched, bandwidth consumption). Instead, it only needs the statistics on the number of viewers of each stream (viewership) in each user group. In this sense, a ``viewer'' in PLVER actually means the corresponding resource demand to each live stream. \nop{Only after such demands are estimated can we design good video replication schedules.}

\nop{Keeping this in mind, the ``assignment'' of viewers to edge servers in Algorithm~\ref{algo:scer} (i.e., $x_{ij}$, $a_i$) implies video allocation to edge servers. It does \textbf{not} imply the real-time request dispatch for a specific viewer. Real-time request dispatch happens after video replications and is out of the scope of the paper. Actually, after video replications, real-time request dispatch is pretty simple: a request is served by the edge directly if the requested video segment has been cached at the edge, otherwise the request is forwarded to the original server. }
\end{remark}

\begin{remark}
We introduce the reward for caching a stream $s$ in a certain edge server $j$ (line~\ref{reward} in Algorithm~\ref{algo:scer}). It implies the traffic demand that could be served by caching the video segments of this stream in server $j$ (during $T$). Let $b$ denote the bitrate of stream $s$; then the reward of $s$ could be defined as following:
\begin{equation}\label{eqt:reward_func}
\mathcal{R}(s,j) = b * \textit{min}\{\lfloor \frac{\Bar{B}_j}{b} \rfloor, N\},
\end{equation}
where $\Bar{B}_j$ is the current available bandwidth of edge server $j$, and $N$ is the number of viewers on stream $s$ that have not been assigned to a server (i.e., $a_i = \varnothing$). The reward is in accord with our objective~(\ref{opt_target}), i.e., maximizing the amount of traffic served by edge servers. \nop{By calculating the reward of each stream, \textit{PLVER} can always pick the stream with the highest value to cache in each round.}
\end{remark}

\setlength{\textfloatsep}{18pt}
\section{Experimental Setup}\label{sec:exp_setup}

\subsection{Live Video Viewership Dataset}\label{sec:TwitchData}
Twitch provides developers with a RESTful API to obtain the live video information. In our experiment, we use a public dataset~\cite{live} that consists of the traces of thousands of live streaming sessions on Twitch~\cite{pires2015youtube}. The dataset contains the information of all live channels in the Twitch system, with a sampling interval of $5$ minutes. Detailed information includes the number of viewers of each channel, bitrates of each channel, and the duration of live sessions. We select the live channels that have more than $100$ viewers and extract the required information of these channels.

Fig.~\ref{fig:nViewers} shows the total number of viewers in the system from Jan. 06 to Jan. 09. During a certain time period, a channel can be either \textit{online}, which means that it is broadcasting a live video, or offline. When a channel is online, we say that it corresponds to a \textit{session}. Fig.~\ref{fig:viewers_dist} shows the distribution of sessions with different average number of viewers.

Fig.~\ref{fig:bitrateCdf} illustrates the distribution of bitrates of channels in the dataset. Based on the video encoding guidelines~\cite{youtubencode}, we assume that the video streams can be encoded with multiple standard resolutions (or bitrates): $240$p, $360$p, $480$p and $720$p (or $400, 750, 1000, 2500$ \textit{Kbps}). Obviously, while a channel broadcasts with bitrate $b$, the viewers of this channel cannot select the video quality with a bitrate exceeding $b$.

\nop{For a channel with the broadcast bitrate $b$, we assume that the arrivals of viewers requesting any broadcast quality follow the Poisson process with an arrival rate $\lambda_i$, where $i$ is the index of $b$ in the list of standard video quality.}

\nop{For the viewers of each channel, we assume that their arrivals (i.e., entering into the channel) and the videos they watch follow the Poisson process. To be specific, for a channel with the broadcast bitrate $b$, the arrivals of viewers requesting any broadcast quality follow the Poisson process with an arrival rate $\lambda_i$, where $i$ is the index of $b$ in the list of standard video quality.}

\subsection{Target Network \& User Groups}\label{setup:ugroup}
geoISP~\cite{geoISP} collected the detailed performance and region coverage information of $2,317$ Internet Service Providers (ISPs) in the US. Based on the information, we build a target network over two US states (Washington and Oregon). We further develop a web crawler to collect the ISP coverage information of $470$ cities over $70$ counties in the two states from the website of geoISP.

We divide all the viewers from these two states (about $0.3$ million on average) into $1253$ user groups (based on the combination of ISP and city) within our target network. Note that one ISP can cover multiple cities and one city can be covered by multiple ISPs. From the dataset, we know the percentage of users in a city that is supported by a particular ISP. For each live stream, we distribute its viewers among these user groups based on the population of each user group (calculated based on each city's population and the ISP coverage percentage of the city).

\subsection{Edge Server Clusters}\label{setup:edgeCluster}

\subsubsection{Setup of edge clusters \& servers} 

\begin{table}[!t]
    \centering
    \setlength{\arrayrulewidth}{0.25mm}
        \caption{Different levels of preferred edge cluster by user groups.}\label{tbl:alloation_define}
    \renewcommand{\arraystretch}{1.03}
    \begin{tabular}{| m{2.4cm}<{\centering} | m{5.6cm}|}
    \hline
    \textbf{Preference Priority} & \textbf{Clusters Description} \\
    \hline
    \textit{Lv. 1}  & clusters that are within the same ISP and located in the same city. \\
    \hline
    \textit{Lv. 2} & clusters that are within the same ISP and located in the same county. \\
    \hline
    \textit{Lv. 3} & clusters that are located in the same city while with different ISPs.\\
    \hline
    \textit{Lv. 4} & clusters that are within the same ISP and located in the same state.\\
    \hline
    \textit{Lv. 5} & clusters that are located in the same county while with different ISPs.\\
    \hline
    \textit{Lv. 6} & clusters that are located in the same state while with different ISPs.\\
    \hline
    \end{tabular}
\vspace{-0.1in}
\end{table}

Among all user groups, we further extract $641$ city-ISP combinations as the target for deploying the edge clusters. Each edge cluster in our experiments consists of five types of edge servers with $5$, $10$, $20$, $40$ and $80$ Mpbs bandwidth capacity, respectively. The servers are randomly deployed at each edge cluster. The total bandwidth of all the deployed edge clusters is set to equal the total traffic demand of all viewers. Note that such bandwidth setting may not always guarantee the full satisfaction of all viewers' demand because the bandwidth of each edge cluster may not be fully utilized and also because the cache capacity and QoE of each edge cluster may be different. Nevertheless, our later experiment shows that such a setting is appropriate to evaluate the performance of different edge caching strategies.

\nop{
it is consistent with our optimization target (given by~(\ref{opt_target})) and thus well-suited to evaluate the performances of different edge caching strategies.}

\nop{
The tremendous amount of edge servers should be distributed among these user groups as well. Among the $1253$ user groups in our target network, we further selected $641$ user groups (where there are more than $50$ local viewers) to deploy our edge servers. Without loss of generality, we assume that there are globally five different types of edge servers which could provide $5$, $10$, $20$, $40$ and $80$ Mpbs bandwidth capacity respectively. We randomly deploy the five types of edge servers into the $641$ user groups such that edge server clusters could be formed in each user groups. 

The deployment of edge servers is conducted repeatedly until the total bandwidth of edge servers is equal to the total traffic demand from viewers. Due to the existence of QoE and cache capacity constraints, this setting cannot guarantee that all viewers' demand could be fully satisfied. Nevertheless, it is well-suited for us to evaluate the performance of edge caching algorithms since it is consistent with our optimization target~(\ref{opt_target}).
}

\subsubsection{Setting of cache capacity} 
For an edge server with bandwidth capacity $b$ Mbps, it should have at least $b * T$ Mb ($T$ is the considered time period) cache capacity to ensure that it has enough resources in our edge replication strategy. For simplicity, we use $\hat{b}$ to denote $b*T$ hereafter. Since video traffic delivery is network intensive, cache capacity of edge servers is normally larger than $\hat{b}$ Mb. In our experiments, we assume that the cache capacity (variable $X$) of all edge servers is uniformed distributed within the range of $(0.5 * \hat{b}, 2 * \hat{b})$. In the following section, we will further adjust the capacity that could be used in each edge server by setting different values of replication cost constraint factor $\alpha$.

\begin{table}[!t]
    \setlength{\arrayrulewidth}{0.25mm}
        \caption{Allocation results of ISOA.}\label{tbl:alloation_results}
    \renewcommand{\arraystretch}{1.4}
    \centering
    \begin{tabular}{| m{2.5cm}<{\centering} |c c c| }
    \hline
    \multirow{2}{*}{\tabincell{c}{Preference rank of \\the allocated cluster}} & \multicolumn{3}{c|}{\# of user groups at each preference level}\\
    \cline{2-4}
    & Greedy Allocation & ISOA & Changes\\
    \hline
    \textit{Lv.1} & 390 & 451 & +61\\
    \hline
    \textit{Lv.2} & 496 & 457 & -39\\
    \hline
    \textit{Lv.3} & 120 & 106 & -14\\
    \hline
   \textit{Lv.4} & 136 & 127 & -9\\
    \hline
    \textit{Lv.5} & 56 & 57 & +1\\
    \hline
    \textit{Lv.6} & 36 & 37 & +1\\
    \hline
    un-allocated & 19 & 18 & -1\\
    \hline
    \end{tabular}
\end{table}

\begin{figure}
    \centering
    \includegraphics[width=0.55\columnwidth]{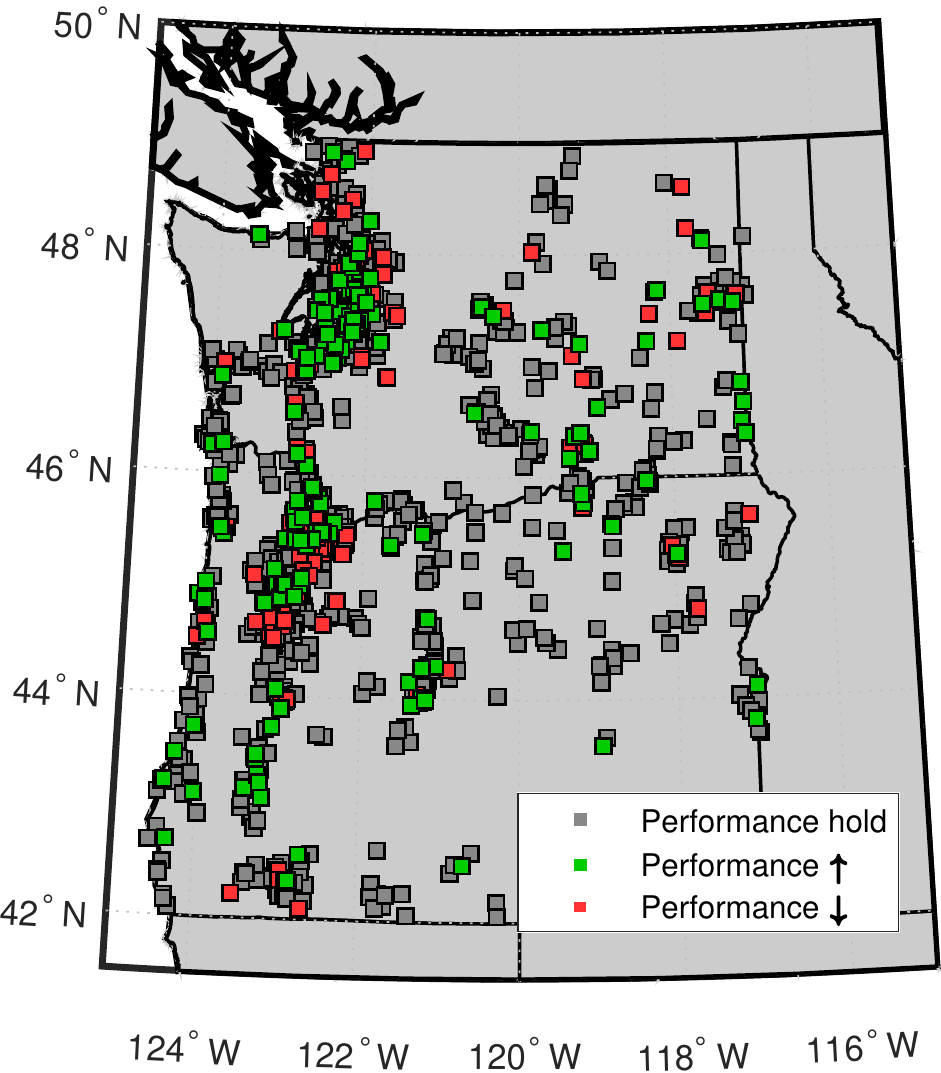}
    \vspace{-0.1in}
    \caption{Performance change with ISOA over \textit{greedy allocation}.}
    \label{fig:stableAllocation_map}
   \vspace{-0.15in}
\end{figure}

\begin{figure*}[!t]
	\centering
	\hspace{-1in}
	\subfigure[Avg. traffic offloading ratio with different $\alpha$.]{
		\label{fig:overall_performance}
		\includegraphics[width=0.32\textwidth]{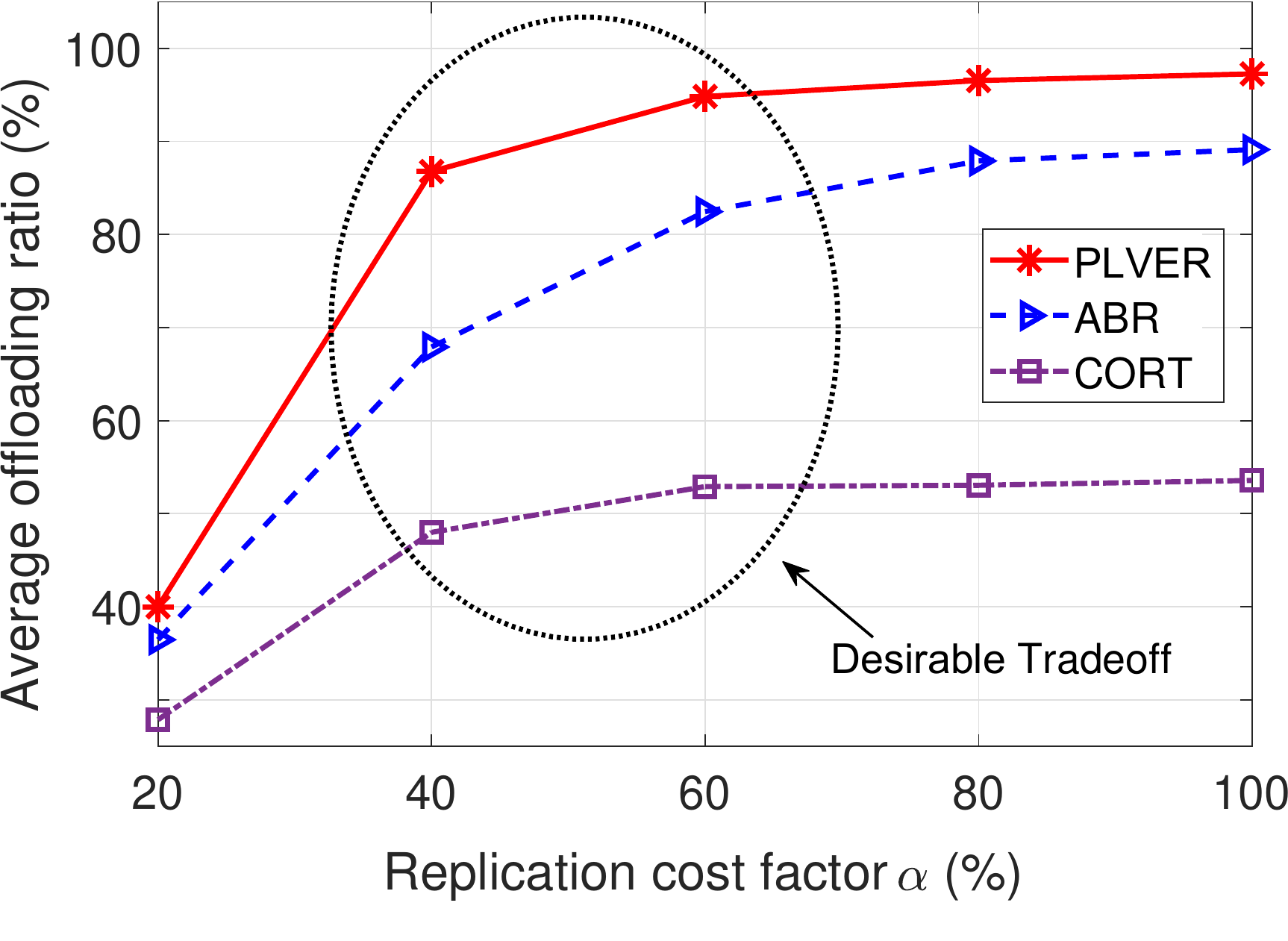}}
	\subfigure[The hourly performance of \textit{PLVER} and ABR.]{
		\label{fig:seer_perform_time}
		\includegraphics[width=0.66\textwidth]{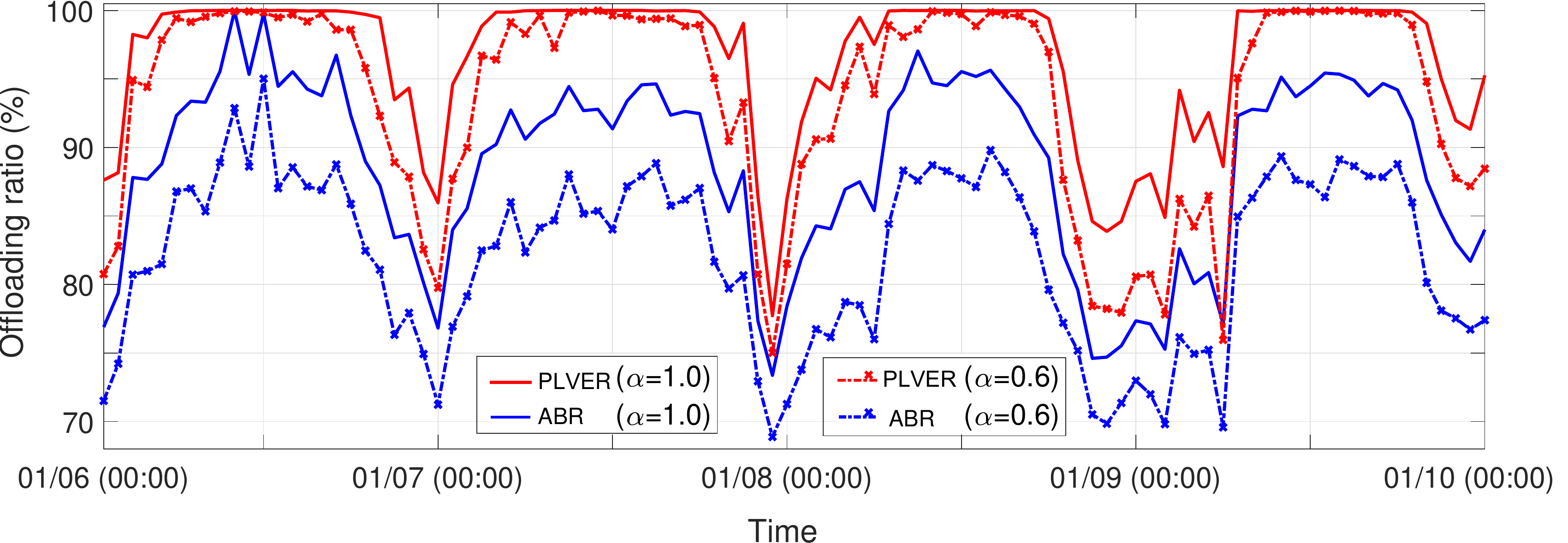}}
	\hspace{-1.05in}
	\vspace{-0.1in}
	\caption{Experiment results of performance of \textit{PLVER} and other methods.}
	\label{fig:PER_exp1}
\end{figure*}

\section{Performance Evaluation}\label{sec:evaluation}

\subsection{Performance Evaluation of Stable One-to-multiple Allocation}

\subsubsection{Evaluation Methodology}

We compare the performance of \textit{ISOA} with another edge cluster allocation strategy: \textit{greedy allocation}. With the greedy allocation, each of the user groups selects its most preferred edge cluster iteratively, until all user groups get allocated or there is no available edge cluster.

\subsubsection{Preference List Generation}
We define the rank of preferred edge clusters of user groups (introduced in~\cref{sec:stable_allocation}), as listed in Table~\ref{tbl:alloation_define}. Note that the preference list defined in Table~\ref{tbl:alloation_define} is just an example paradigm to generate the input of our stable allocation algorithm, and it can be altered by the CDNs themselves, e.g., according to the contract terms under which the cluster is deployed, granularity of the user groups partition, and so on~\cite{maggs2015algorithmic}.

\subsubsection{Performance Evaluation of ISOA}

We conduct the stable one-to-multiple allocation between user groups and edge clusters based on the data introduced in~\ref{setup:ugroup} and \ref{setup:edgeCluster}, using \textit{ISOA} and the aforementioned \textit{greedy allocation} method. The detailed allocation results are summarized in Table~\ref{tbl:alloation_results}. Since there are $1253$ user groups in our experiment, the table shows the distribution of these user groups being allocated with different levels of preferred edge cluster. For example, there are $390$ user groups that are allocated with their first ranked (most preferred) edge cluster with \textit{greedy allocation}, while the number is increased to $451$ with \textit{ISOA}. Compared with the greedy allocation, \textit{ISOA} can allocate more user groups with their higher ranked (more preferred) edge clusters. The performance improvement with \textit{ISOA} over the greedy allocation for every user groups in our experiment are illustrated in Fig.~\ref{fig:stableAllocation_map}, where the performance of each user group is marked with a colored square.
\vspace{-0.1in}
\subsection{Performance Evaluation of Proactive Edge Replication}

\subsubsection{Evaluation Methodology}

We evaluate \textit{PLVER} by comparing it with the following replication strategies:

\begin{itemize}
    \item \textit{Auction Based Replication (ABR):} Each edge server conducts a simple ``auction'' to determine the cached videos: live videos with the largest number of viewers via the edge server win the auction and are cached, and the auction repeats until the edge server uses up its cache capacity~\cite{hung2018combinatorial}. In other words, this method replicates the videos into an edge server based on their current number of viewers in the decreasing order.
    \item \textit{Caching On Requested Time (CORT):} \textit{This strategy does NOT adopt pre-replication and using request triggered caching strategy instead.} It caches the videos into the edge servers in real-time when video segments are truly requested by the end users. When content requested, it first checks if there are available edge servers to serve this request; if not, it replicate the video segments of this stream into a new edge server.
\end{itemize}

\nop{We implement different replication strategies and replicate the newly generated live video segments into different edge servers accordingly. To be specific, we conducted real-time request redirection to redirect live video access requests to the edge servers. Note that this viewership data are derived from the next time window of length $T$.\nop{ for which we have generated the replication schedule.  we used to generate the guidance.} Also, the requests that cannot be responded by any cached edge server will be directed to the back-end cloud. }

To evaluate the performance of different strategies, we use the metric \textbf{\textit{offloading ratio}}, which is calculated by the amount traffic served by the edge servers divided by that of the overall traffic in the time period of length $T$. The performance is evaluated under different values of \textbf{\textit{replication cost factor $\alpha$}} (refer to~\cref{subsec:replicationCost}), so that we can investigate the tradeoff between performance and replication overhead.

\subsubsection{Overall Performance of PLVER}

Based on the twitch viewership data from Jan. 06, 2014 to Jan. 09, 2014, we conduct experiments on an hourly base. By setting the value of $\alpha$ to $20\%, 40\%, 60\%, 80\%$ and $100\%$, we compute the average offloading ratios of the three strategies in each case. The results  are shown in Fig.~\ref{fig:overall_performance}, from which we can see that \textit{PLVER} outperforms \textit{ABR} and \textit{CORT} in all five cases. The overall performance improvement by \textit{PLVER} for the five cases are $9\%, 10\%, 15\%, 28\%$ and $10\%$ over \textit{ABR}, respectively, and $82\%, 82\%, 79\%, 81\%$ and $44\%$ over \textit{CORT}, respectively. 

Furthermore, we can find from Fig.~\ref{fig:overall_performance} that the overall performance got a considerable improvement when the replication cost constraint ($\alpha$) is increased from $20\%$ to $60\%$. However, the performance improvement fades when $\alpha$ continues to increase after $60\%$. This situation holds for all the three replication strategies. Therefore, in our experiment, it reaches a good tradeoff between performance and replication costs when $\alpha$ is between $40\%$ and $60\%$ (as shown in Fig.~\ref{fig:overall_performance}).

\subsubsection{Detailed Performance of PLVER}

\begin{figure}[!t]
    \vspace{-0.06in}
    \centering
    \includegraphics[width=1\columnwidth]{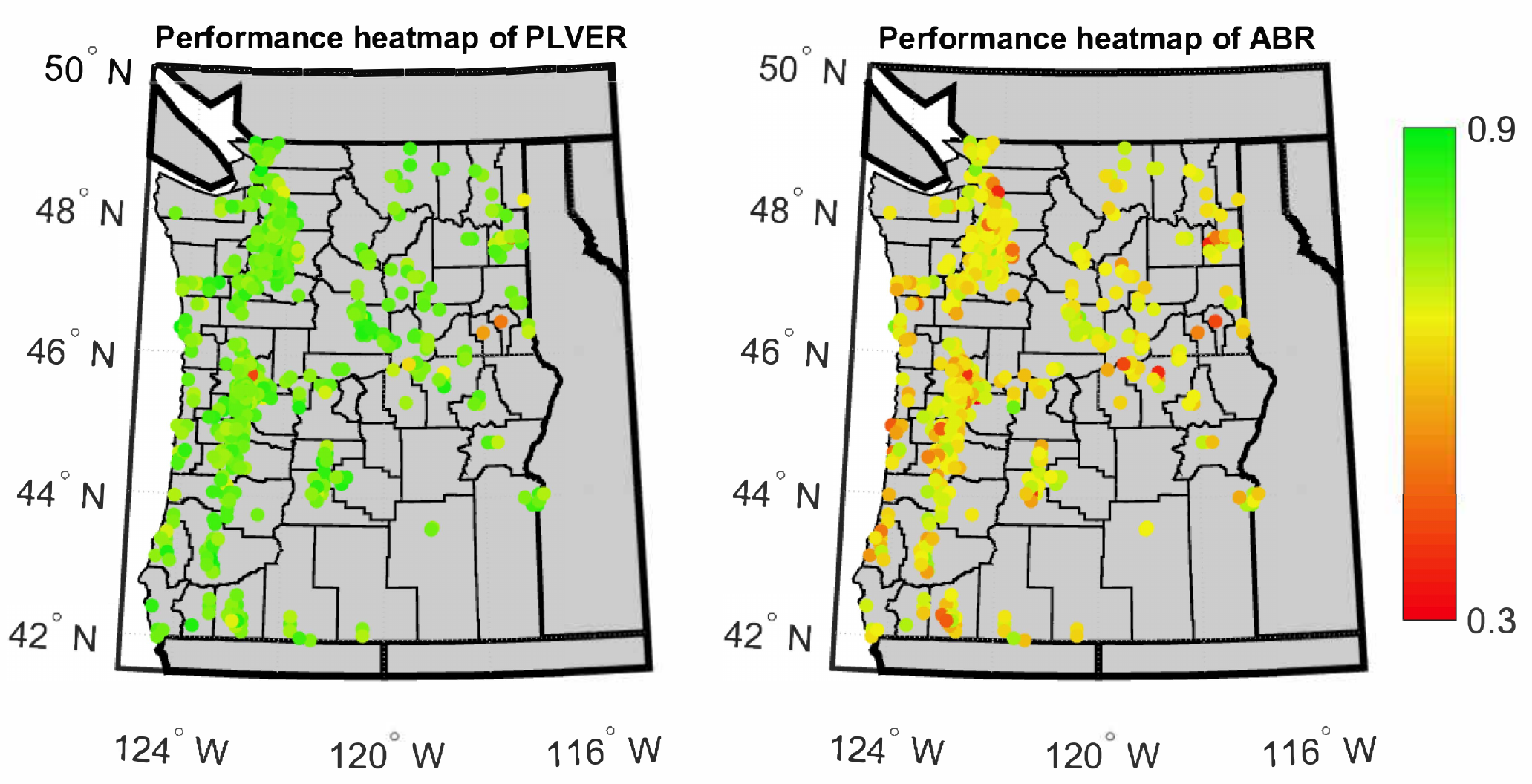}
    \vspace{-0.115in}
    \caption{The performance of \textit{PLVER} and ABR in each edge cluster with $\alpha = 0.4$.}
    \label{fig:p_heatmap}
    \vspace{-0.05in}
\end{figure}

Referring to the overall performance, \textit{ABR} is more comparable to \textit{PLVER} (than \textit{CORT}). We thus investigate the detailed performance behaviors of \textit{PLVER} and \textit{ABR}. The traffic offloading ratios for i) each hour and ii) each user group are shown in Fig.~\ref{fig:seer_perform_time} and Fig.~\ref{fig:p_heatmap}, respectively.

Fig.~\ref{fig:seer_perform_time} shows the hourly traffic offloading ratio of \textit{PLVER} and \textit{ABR} with the replication cost constraint factor $\alpha$ equal to $100\%$ and $60\%$, respectively. It shows that even within non-peak hours (when the resources of edge servers are sufficient), it is hard for \textit{ABR} to yield a satisfied performance. In contrast, when $\alpha$ decreases from $100\%$ to $60\%$, the performance degradation of \textit{PLVER} is much smaller than that of \textit{ABR}.

Fig.~\ref{fig:p_heatmap} shows a heat map indicating the performance of \textit{PLVER} and \textit{ABR} at each edge cluster (with $\alpha = 40\%$), where the traffic offloading ratios are represented by different colors.  Since there are no edge clusters with performance less than $30\%$ or greater than $90\%$, our color bar denotes the traffic offloading ratio from $30\%$ to $90\%$. An edge cluster with better performance is colored in green, and worse in red.

We also investigate the performance when requesting different video qualities ($240$p, $360$p, $480$p, $720$p). The satisfaction ratio of requests (i.e., the ratio of requests that are successfully directed to corresponding edge servers) with different replication strategies are shown in Fig.~\ref{fig:request_success_bitrate}. We can observe that \textit{PLVER} outperforms \textit{ABR} and \textit{CORT} for all types of quality requests. Among the four different quality requests, the high quality request of $720$p is with relatively low traffic offloading ratio than that of the other three video qualities. However, as high quality requests generate more traffic than the others, it impacts more on the final performance. \textit{PLVER} provides a satisfaction ratio of $36\%$ for the $720$p requests, which is higher than those of \textit{ABR} ($19\%$) and \textit{CORT} ($30\%$), respectively.

\subsubsection{Impact of Viewership Fluctuation}

As \textit{PLVER} makes use of the viewership information (i.e., the number of viewers) in current time window to make decisions in next time window, the viewership fluctuation in consecutive time slots may impact the performance of replication algorithms. To investigate that, we first generate the replication schedules by different replication strategies referring to the viewership data in peak traffic hours of~\cref{sec:TwitchData}, then we manually generate a new viewership data to test the performance of these replication schedules. The new viewership data is generated by introducing different levels of fluctuations on the former viewership data that we used to generate the replication schedules. To be more specific, the number of viewers of each channel are added with different percentages of fluctuation (e.g., randomly plus or minus $20\%$).

\begin{figure}[!t]
    \vspace{0.05in}
	\centering
	\hspace{-1.05in}
    \subfigure[The satisfaction ratio for video requests with different qualities.]{
        \label{fig:request_success_bitrate}
        \includegraphics[width=0.48\columnwidth]{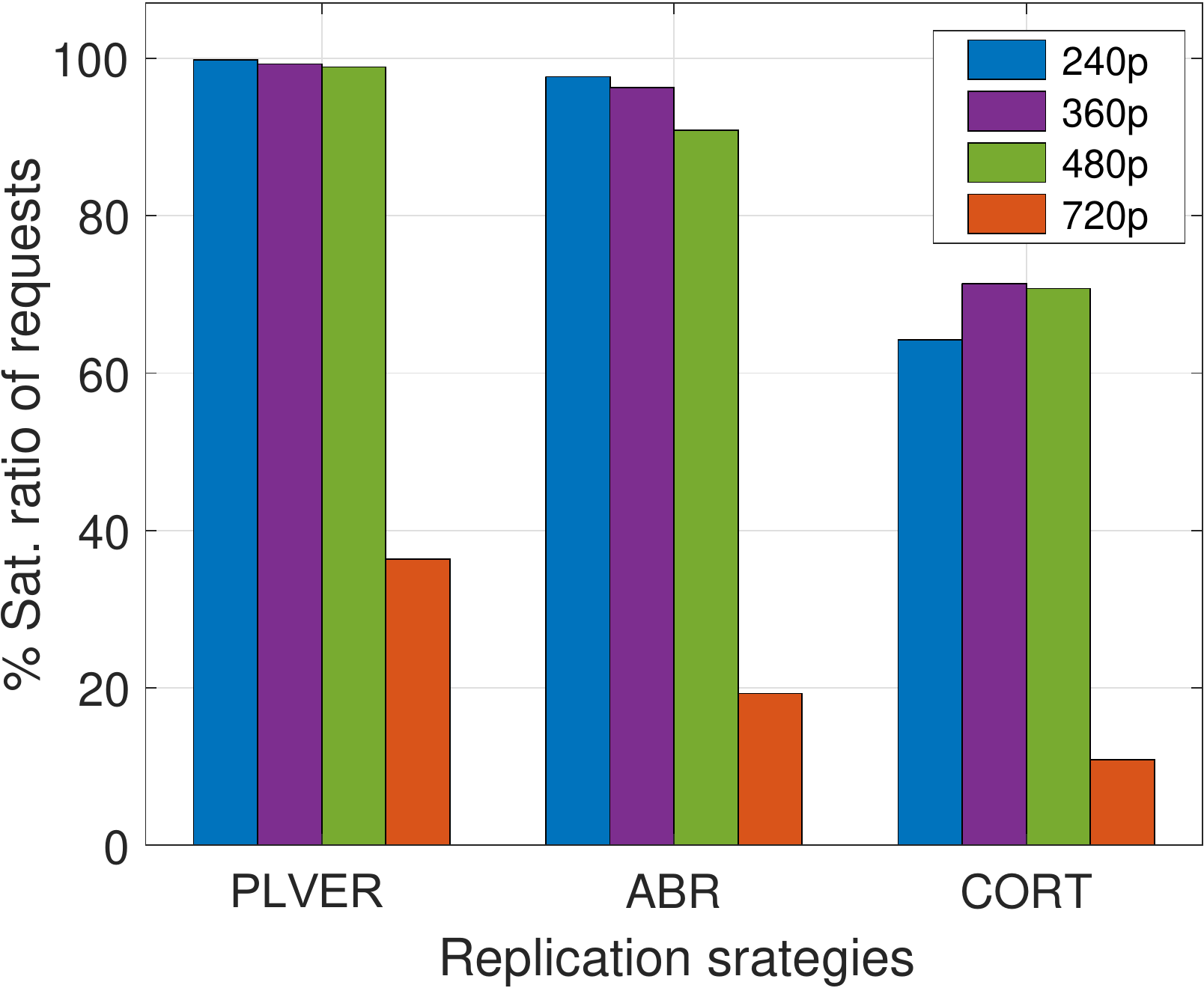}
    }
    \subfigure[The variation of performance with the fluctuation change on number of stream viewers.]{
        \label{fig:perform_on_fluct}
        \includegraphics[width=0.48\columnwidth]{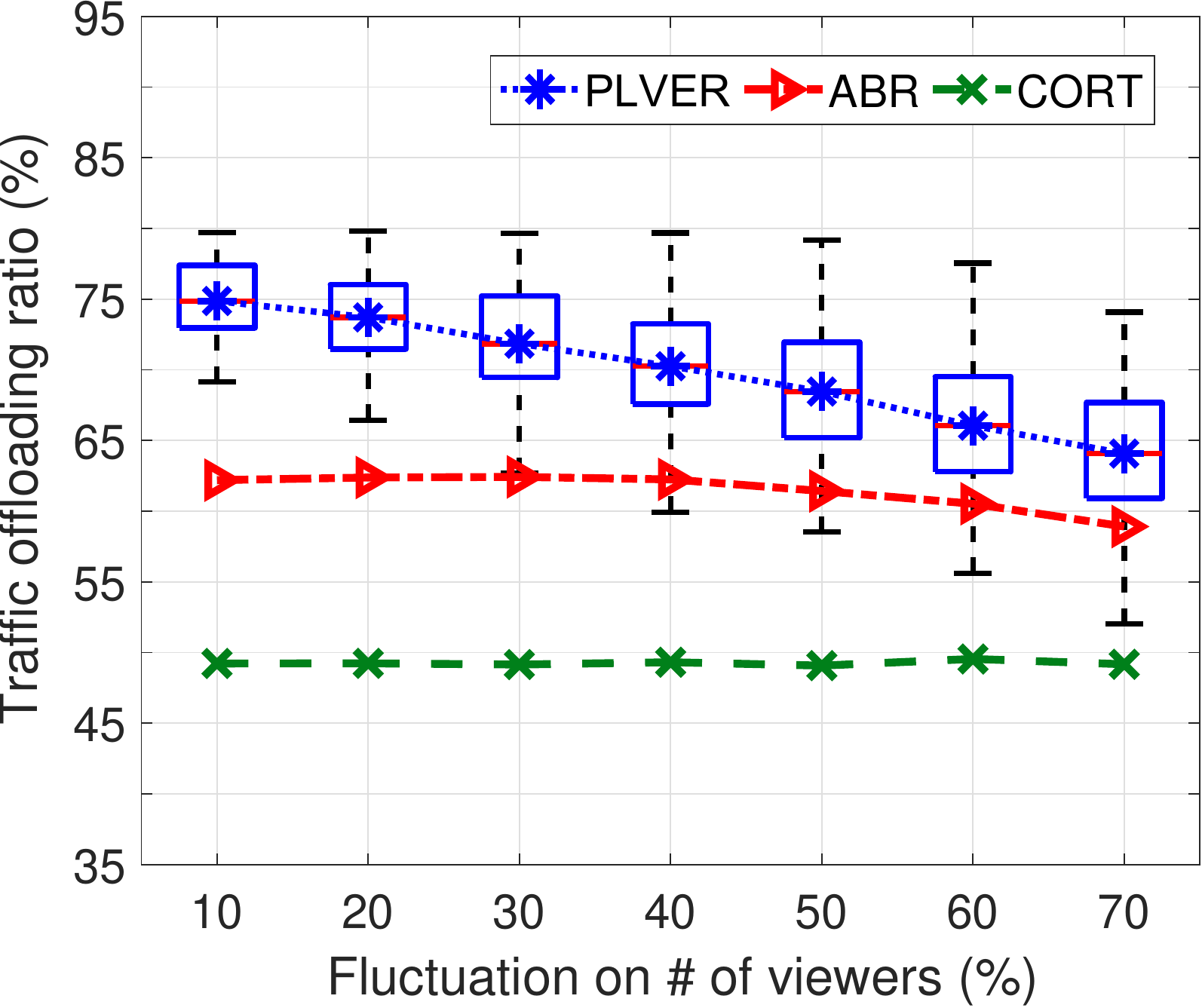}
    }
	\hspace{-1.05in}
	\vspace{-0.08in}
	\caption{Performance results of \textit{PLVER} and other replication strategies.}
	\vspace{-0.03in}
	\label{fig:PER_exp2}
\end{figure}

The performance of \textit{PLVER} under different viewership fluctuations is shown in Fig.~\ref{fig:perform_on_fluct}. We can see that the performance curve (representing the traffic offloading ratios) slightly goes down from $75\%$ to $64\%$ with fluctuations changing from $10\%$ to $70\%$. Nevertheless, according to the statistical analysis of our dataset, viewership fluctuations higher than $30\%$ are quite rare.  Hence, its impact on \textit{PLVER} is quite small.

\section{Conclusion}
Live video services have gain extreme popularity in recent years. The QoE of live videos, however, suffers from the cache miss problem occured in the edge layer. Solutions from the current live video products as well as the state-of-the-art researches would pose extra latency to the live streams which sacrifices the ``liveness'' of delivered video. In this paper, we propose \textit{PLVER}, an efficient edge-assisted live video delivery scheme aiming at improving the QoE of live videos. \textit{PLVEr} first conducts a \textit{one-to-multiple} stable allocation between edge clusters and user groups. Then it adopts proactive video replication algorithms over the edge servers to speed up the video replication over edge servers. Trace-driven experimental results demonstrate that our solution outperforms other edge replication methods.

\bibliographystyle{IEEEtran}
\bibliography{reference}

\begin{IEEEbiography}[{\includegraphics[width=1in,height=1.25in,clip,keepaspectratio]{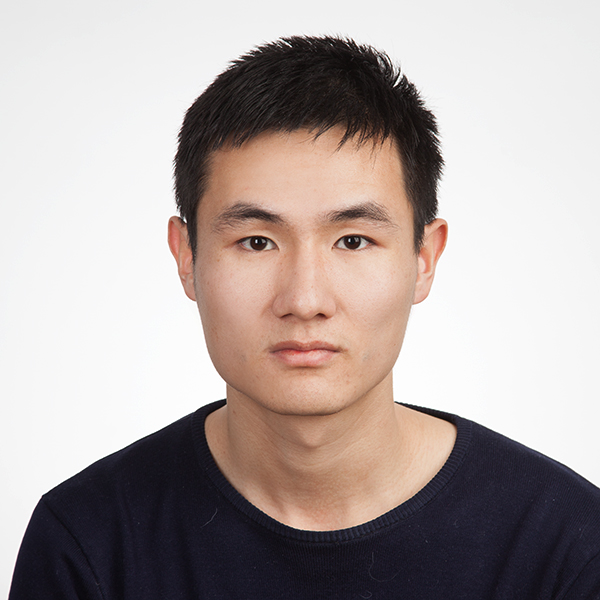}}]{Huan Wang}
received the Bachelor and Master degrees in computer science from Southwest Jiaotong University and the University of Electronic Science and
Technology of China, in 2013 and 2016, respectively. He is currently pursuing Ph.D. degree with the Department of Computer Science, University
of Victoria, BC, Canada. His research interests include content/video delivery, edge caching and computing, network traffic anomaly detection.
\end{IEEEbiography}

\begin{IEEEbiography}[{\includegraphics[width=1in,height=1.25in,clip,keepaspectratio]{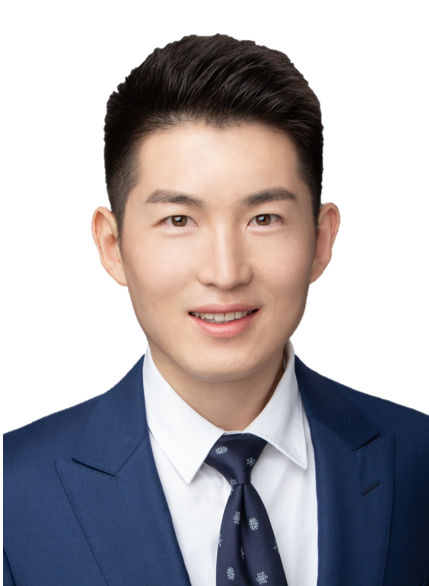}}]{Guoming Tang}
(S'12-M'17) is currently a research fellow at the Peng Cheng Laboratory, Shenzhen, Guangdong, China. He received his Ph.D. degree in Computer Science from the University of Victoria, Canada, in 2017, and the Bachelor's and Master's degrees from the National University of Defense Technology, China, in 2010 and 2012, respectively. He was also a visiting research scholar of the University of Waterloo, Canada, in 2016. His research mainly focuses on cloud/edge computing, green computing, and intelligent transportation systems.
\end{IEEEbiography}

\begin{IEEEbiography}[{\includegraphics[width=1in,height=1.25in,clip,keepaspectratio]{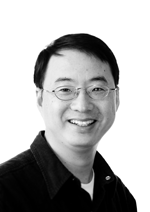}}]{Kui Wu}
(S'98-M'02-SM'07) received the BSc and the MSc degrees in computer science from the Wuhan University, China, in 1990 and 1993, respectively, and the PhD degree in computing science from the University of Alberta, Canada, in 2002. He joined the Department of Computer Science, University of Victoria, Canada, in 2002, where he is currently a Full Professor. His research interests include network performance analysis, mobile and wireless networks, and network performance evaluation. He is a senior member of the IEEE.
\end{IEEEbiography}

\begin{IEEEbiography}[{\includegraphics[width=1in,height=1.25in,clip,keepaspectratio]{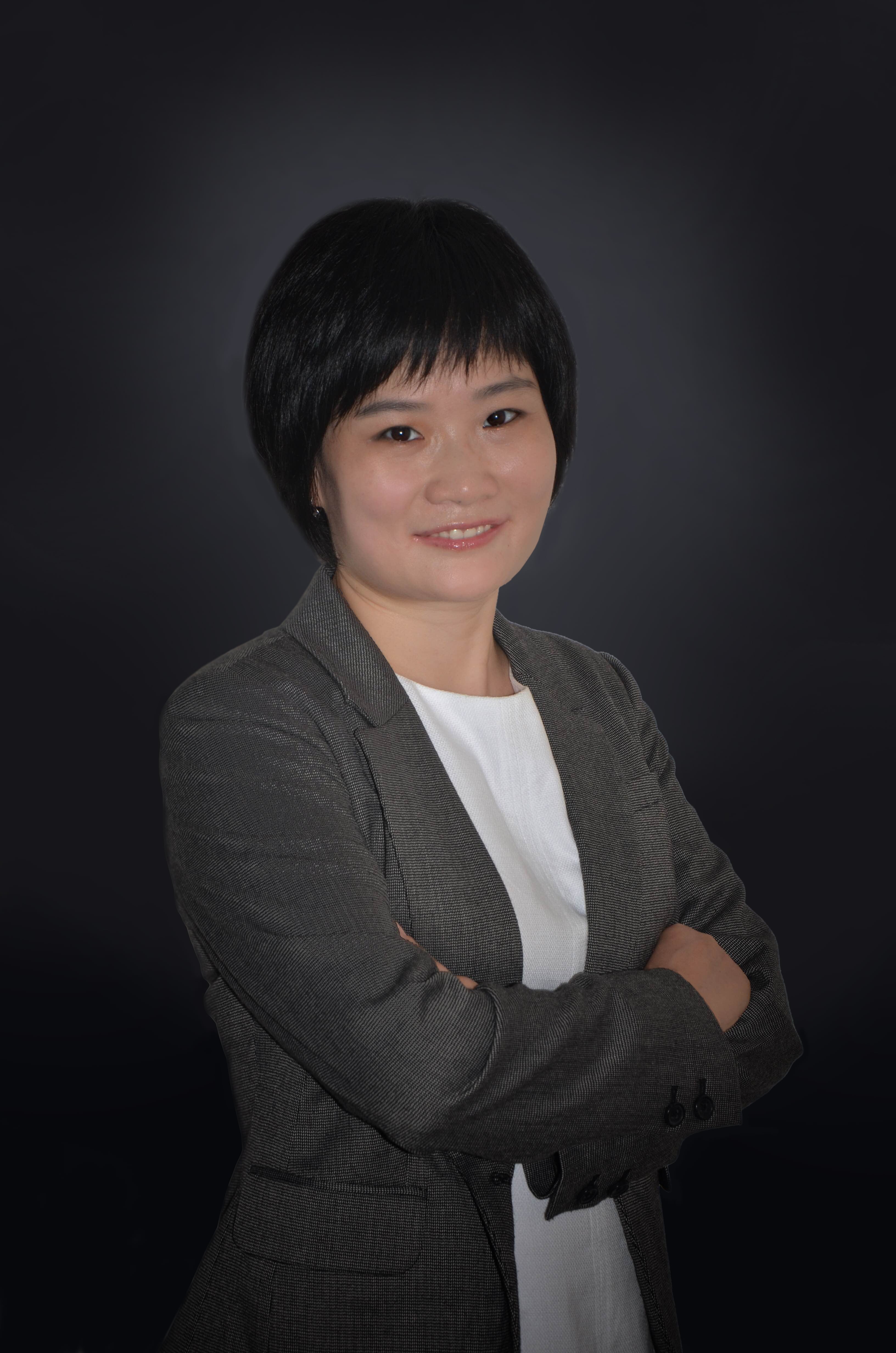}}]{Jianping Wang}
is a professor of the computer science department at City University of Hong Kong, Hong Kong. She received the B.E. degree and MSc degrees in computer science from Nankai University, Tian Jin, China, in 1996 and 1999, respectively and the Ph.D. degree in computer science from University of Texas at Dallas, USA, in 2003. Her research interests include cloud computing, service oriented networking, and data center networks.
\end{IEEEbiography}

\end{document}